\documentstyle[prd,aps,preprint,tighten,eqsecnum,epsf]{revtex}
\begin{document}
\preprint{TUW-95-18 (revised version)}
\title{New variables, the gravitational action, and\\
boosted quasilocal stress-energy-momentum\footnote{Revised version,
to appear in {\em Classical and Quantum Gravity}.}}
\author{Stephen R. Lau\footnote{Previously at the
Institute of Field Physics,
Department of Physics \& Astronomy,
University of North Carolina,
Chapel Hill, NC 27599--3255 USA.}}
\address{Institut f\"{u}r Theoretische Physik, 
Technische Universit\"{a}t Wien\\
Wiedner Hauptstra{\ss}e 8-10,
A-1040 Wien, AUSTRIA\\
email: lau@tph16.tuwien.ac.at\\
{\rm and}\\
Inter--University Centre for
Astronomy and Astrophysics, University of Poona Campus\\
Post Bag 4, Ganeshkhind,
Pune, 411 007 INDIA}
\maketitle

\begin{abstract}
This paper presents a complete set of quasilocal densities which
describe the stress-energy-momentum content of the gravitational
field and which are built with Ashtekar variables. The densities
are defined on a two-surface $B$ which bounds a generic spacelike
hypersurface $\Sigma$ of spacetime. The method used to derive the
set of quasilocal densities is a Hamilton-Jacobi analysis of a
suitable covariant action principle for the Ashtekar variables.
As such, the theory presented here is an Ashtekar-variable
reformulation of the metric theory of quasilocal
stress-energy-momentum originally due to Brown and York.
This work also investigates how the quasilocal densities behave
under generalized boosts, i. e. switches of the $\Sigma$ slice
spanning $B$. It is shown that under such boosts the densities
behave in a manner which is similar to the simple boost law 
for energy-momentum four-vectors in special relativity. 
The developed formalism is used to obtain a collection of
two-surface or boost invariants. With these invariants, one may
``build" several different mass definitions in general relativity, 
such as the Hawking expression. Also discussed in detail in this 
paper is the canonical action principle as applied to bounded 
spacetime regions
with ``sharp corners."
\end{abstract}
Vienna, March 1996
\pacs{???}

\section{Introduction}

The geometric expression for the energy of a nonrelativistic system
(the functional form of the Hamiltonian in terms of the coordinates
and momenta) can be discerned from the system's action functional.
This follows from a basic tenet of Hamilton-Jacobi theory: the
classical energy of the system is minus the rate of change of the
classical action (the Hamilton-Jacobi
principal function) with respect
to a unit stretch in the absolute Newtonian time.
The ability to define
the classical energy in this way rests on the fact that in the
conventional variational principle for the system the
lapse of absolute time is fixed as boundary data. From a
practical
standpoint, this means that in order to find the
geometric expression
for the system's Hamiltonian one need only consider the
general variation of the action in
which the endpoints of trajectories in the variational
set are not
held fixed (known as the Weiss action principle \cite{Munkunda}).
Upon inspection of the boundary-term contributions to the variation,
one can determine the canonical momenta
as the factors which multiply
the variations in the endpoint values of the coordinates.
Furthermore, after the momenta are determined,
careful inspection of
the boundary-term factor with multiplies the
variation in the absolute
time then reveals the functional form of the Hamiltonian.

Brown and York have proposed a generalization of
the Hamilton-Jacobi method, which is applicable to a wide class
of generally covariant field theories
of a spacetime metric (in any
dimension); and they have used this generalized method to
discern what geometric expressions play the role of quasilocal
stress, energy, and momentum in general relativity.\cite{BY}
Field theories
of a spacetime metric enjoy a crucial feature in
common with simple
nonrelativistic systems: in the action principle it
is possible to
fix the time as boundary data. To see that this is
indeed the case, consider a spacetime region ${\cal M}$ which is
topologically the Cartesian product of Riemannian
three-manifold $\Sigma$ and a closed connected segment
of the real
line $I$. The three-manifold $\Sigma$ has a boundary
$\partial \Sigma = B$
(which need not be connected). Therefore, one element of the boundary
$\partial {\cal M}$ of ${\cal M}$ is a three-dimensional
timelike hypersurface
$\bar{\cal T}$ (``unbarred" ${\cal T}$ is reserved
for a more special meaning) which has the
topology of $I \times B$ and is a $(2+1)$-dimensional
spacetime in
its own right. The other boundary elements are $t'$, the
three-manifold corresponding to the
initial point of $I$, and $t''$,
the three-manifold corresponding to the final point of
$I$.\footnote{One may imagine that
${\cal M} \subset {\cal U}$, where
$\cal U$ is some ambient spacetime known as the
{\em universe} or sometimes the {\em heat bath}. The boundary $B$
and its history $\bar{\cal T}$ are simply collections
of points in
$\cal U$ and need not be physical barriers.} 
Such a spacetime region is depicted in Figure (\ref{spacetimefig}).
Now suppose
that we are given
a ``suitable" action functional for the metric (and possibly
matter) fields on the spacetime region ${\cal M}$. By ``suitable"
we mean
that the variational principle associated with the
action features
fixation of the induced metric on each
of the boundary elements $t'$, $t''$, and $\bar{\cal T}$.
In particular, the lapse of proper time between the initial and
final hypersurfaces is fixed as boundary data since this information
is encoded in the fixed $\bar{\cal T}$ three-metric. The quasilocal
energy
is then identified as minus the rate of change of the classical
action with respect to a unit stretch  in the proper time
separation between $t'$ and $t''$. (Therefore, inspection of the
boundary-term contributions
to the variation of the action can reveal the geometric
expression for the quasilocal energy. This geometric expression
is obtained by isolating the factor which multiplies
the variation
in the lapse function which controls the proper time separation
between $B$ slices of $\bar{\cal T}$.) However, notice that
the $\bar{\cal T}$
three-metric provides more than just the lapse of proper time
between the initial and final slices, since it contains information
about all possible spacetime intervals on $\bar{\cal T}$.
One is free to consider the rate of change
in the classical action
which corresponds to arbitrary variations in the
$\bar{\cal T}$ boundary
data. A quasilocal surface stress-energy-momentum tensor corresponds
to this freedom. For the most relevant case of general relativity,
the analysis of
Ref. \cite{BY} has demonstrated how this tensor leads to quasilocal
surface densities for {\em energy}, {\em tangential momentum},
and
{\em spatial stress} (all are pointwise tensors defined on $B$)
which describe the
stress-energy-momentum content of the $\Sigma$ matter and
gravitational fields contained within $B$. The theory of quasilocal
stress-energy-momentum originally proposed in Ref.\ \cite{BY} is
currently being extended considerably. One extension has been the
introduction of quasilocal surface densities for
{\em normal momentum} and {\em temporal stress}. The new
developments associated with this extended theory will appear in
an upcoming paper \cite{BYL}, and the results of the present paper
are based heavily on these new developments (though the analysis
here is reasonably well-contained). For a description of the new
developments to be found in Ref. \cite{BYL} and how they relate
to the present paper, see the discussion section at the end
of this work.

This paper uses a Hamilton-Jacobi-type method to derive quasilocal
stress-energy-momentum surface densities which are built with the
Ashtekar gravitational variables \cite{Ashtekar1}, thereby fully 
extending some preliminary results concerning Ashtekar variables 
and quasilocal energy-momentum given in \cite{Lau}. 
Since the the Ashtekar version of general relativity is 
inherently a non-metric formalism, the Hamilton-Jacobi analysis given
by Brown and York has to be modified. However, though the choice 
of Ashtekar variables over Arnowitt-Deser-Misner (ADM) variables
\cite{ADM} leads to non-trivial technical complications and 
some conceptual ones, our overall goal 
is the same as in the metric theory; and the cornerstone of the 
method used here remains a ``suitable" action principle, 
i. e. information about the lapse 
of proper time must be fixed as boundary data. 
Now, the usual covariant formulation of the Ashtekar variables is
based on the well-known chiral action 
independently given by Samuel \cite{Sam}
and Jacobson and Smolin \cite{Smolin}. This is a Palatini action
which features the independent variation of the spacetime self-dual
spin connection and the $SL(2,C)$ soldering form.
Applied to our spacetime region ${\cal M}$, this action
principle does
not feature fixation of metric data on $\bar{\cal T}$,
and hence it is not
well-suited for our purposes.
Perhaps, one could consider adding the necessary boundary terms
to the chiral action in order to obtain a suitable variational
principle. However, here we follow another route which is based
on a lesser-known covariant
formulation of the Ashtekar theory which has been given by
Goldberg.\cite{Goldberg} Goldberg's action functional is
first-order, but in the variational principle the connection is not
varied independently from the tetrad. We find that, subject to
certain gauge fixation of the tetrad, Goldberg's action is a tetrad
version of the action functional used
to derive quasilocal stress-energy-momentum in the metric scenario.
It should be mentioned now that partial gauge fixation of the
tetrad and triad plays a crucial role in what follows. At first
sight this may seem objectionable. But one should
recall that such gauge fixation is also unavoidable in the triad
formulation of Hamiltonian gravity when one
discusses the notions
of total energy and momentum in the asymptotically-flat scenario.
In that case one must deal with a ``fiducial triad at
infinity."\cite{Goldberg,Ashtekar1} The gauge fixation of the
triad in the quasilocal context is of the same nature.

There is a subtle interpretational issue concerning the analysis
to follow which deserves so comment at the outset. The Brown-York
quasilocal densities
are not unique, since one has the freedom to add a
{\em subtraction term} $- S^{\it 0}$ 
(a functional of the fixed boundary data)
to the gravitational action $S^{\it 1}$ which
is used to derive the densities. Brown and York have offered the
interpretation that such freedom allows one to set the reference
points for the quasilocal densities.\footnote{This is quite
analogous to the situation in nonrelativistic
mechanics, where one can affect the definition of a system's
energy and canonical momenta by adding boundary terms to a
system's action.} Now, the results of gravitational thermodynamics
are, in fact, independent of the choice of subtraction term, and,
therefore, such freedom seems to be an
unnecessary one when examining the statistical mechanics of the
strong gravitational field.\cite{BMYW,micro} However, the
subtraction term plays an important role in
several other theoretical contexts. For instance, it must be
incorporated into the definition of the quasilocal energy, if in
the suitable limit the definition  is to agree with the
ADM notion of energy
at spacelike infinity.\cite{ADM,BY} Furthermore, recent 
research has suggested that
there is an implicit reference point set in some spinor
constructions of gravitational energy based on the Witten-Nester
integral.\cite{Lau4}
In this paper the passage from the triad ADM variables to the
Ashtekar variables is effected by the
addition of a purely imaginary boundary term to the action.
We formally treat this boundary term as a subtraction term 
$- S^{\it 0}$ \`{a}
la Brown and York. This allows us to construct the theory in a
parallel fashion with the presentations given in Refs.
\cite{BY,BYL}. However, though technically this
viewpoint is completely satisfactory, it should be realized that
it is less satisfactory from an interpretational standpoint.
Indeed, if we wish to adopt
the Brown-York interpretation for the imaginary subtraction term,
then we are confronted with the issue of imaginary reference
points for the quasilocal densities. Furthermore, even with the
imaginary subtraction term, in the suitable limit the
Ashtekar-variable expression for the quasilocal energy as given
here does not agree with the ADM notion of energy at spacelike
infinity. This seems alarming, but in fact is not a real problem.
It merely signifies that it is perhaps better to view the
imaginary subtraction term not as a true subtraction term, but
rather as part of a different base action $S'^{\it 1} = S^{\it 1}
- S^{\it 0}$ suitable for the Ashtekar variables. To
derive an expression for the quasilocal energy in terms of the
Ashtekar variables which is in agreement with the ADM expression,
we would need to
begin with an action which differs from this different choice
for the base action by {\em yet another} subtraction term 
$- S'^{\it 0}$ (so the full action would be 
$S' = S'^{\it 1} - S'^{\it 0}$). 
In other words, our analysis is actually performed only on a 
base action $S'^{\it 1}$ (even though we split this base action 
into two pieces and treat one piece formally as a Brown-York 
subtraction term), and it should be understood that in some 
contexts it may be necessary to consider the addition of appropriate 
subtraction terms to this base action. We discuss these issues in 
more detail in the concluding section.

The organization of this paper is as follows. In $\S$ II, the
preliminary section, we discuss in detail the
geometry of ${\cal M}$ in
terms of several classes of spacetime foliations. This discussion
is the groundwork for
the analysis in the main sections. We also collect some notations
and conventions in this section. In $\S$ III we derive a full set
of quasilocal densities which are expressed in terms of the Sen
connection and triad on $\Sigma$, and thus may be easily
rewritten
later on in terms of the canonical Ashtekar variables. The
geometric forms of these densities
are discerned from a careful analysis of the boundary terms
which appear in the Goldberg action principle. This analysis is
quite analogous to the method described for nonrelativistic systems
in the introductory
paragraph. In $\S$ IV we turn to the issue of how the collection of
quasilocal densities behave under generalized boosts.
This behavior is similar to the simple boost law for energy-momentum four-vectors 
in special relativity. With the derived boost relations we then 
show how
to obtain a number of two-surface or boost invariants, one of which
is the Hawking mass.\footnote{In this paper we make a sharp distinction 
between the notion of quasilocal {\em energy}, which is slice-dependent
(observer-dependent) and unique only up to a reference point, and 
notions of quasilocal {\em mass}, which are boost-invariant and uniquely
defined. However, both the quasilocal energy and the notions of quasilocal 
mass considered in this paper depend solely on gravitational 
Cauchy data.}\cite{Hawking}
Also in $\S$ IV, we consider the canonical
form of the action principle for spacetime regions with
``sharp corners." This analysis supplements recent results from
standard metric gravity for such
spacetimes.\cite{BYL,Hayward} The
appendices provide some kinematical results
necessary for the central discussions. The first three appendices
develop the results necessary to write down the boost
relations for the quasilocal densities.
A forth and final appendix presents a
method for dealing with ``corner" terms in gravitational actions
(such terms are described below).

\section{Preliminaries}

\subsection{Foliations} The boundary structure of
${\cal M}$ leads to two
classes of spacetime foliations.
Our discussion of these foliations
is close to one given by Hayward and Wong \cite{Hayward}.

{\em Temporal foliations of ${\cal M}$}.
The first type of break-up
stems from a conventional ADM foliation
of ${\cal M}$ into a family
spacelike hypersurfaces.\ \cite{ADM}
A foliation of this class,
referred to as a {\em temporal foliation}, is specified
by a time function $t:
{\cal M} \rightarrow I$. The leaves of the
foliation or {\em slices} are the
level hypersurfaces of this time coordinate $x^{0} \equiv t$.
Often, the
possible time functions are restricted by the
requirement that both
$t'$ and $t''$ must be level hypersurfaces of
coordinate time.
The letter $\Sigma$ is used both to denote a
foliation of ${\cal M}$ and to refer to
a generic slice of this foliation, and the
$\Sigma$ slice specified by
$t = t_{*}$ ($t_{*}$ is some constant) is denoted $\Sigma_{t_{*}}$.
If the manifolds $t'$ and $t''$ are level hypersurfaces of
coordinate time, then it is convenient to
set $t' = \Sigma_{t'}$ and $t'' = \Sigma_{t''}$.
The timelike, future-pointing, unit,
hypersurface normal of a $\Sigma$ foliation is denoted by $u$.

{\em Radial foliations of ${\cal M}$}.
The existence of the timelike
boundary $\bar{\cal T}$ suggests an
alternative class of foliations
of ${\cal M}$. Members of this alternative
class are called {\em radial
foliations} and rely on {\em timelike} hypersurfaces or
{\em sheets} which have the topology of $\bar{\cal T}$ (informally,
sheets are radial leaves while slices are temporal leaves).  One
assumes that a radial coordinate $x^{3} \equiv r$ parameterizes a
nested family of such hypersurfaces which extend inward from
$\bar{\cal T}$ (see Figure (\ref{radialfoliationfig})). 
This family of timelike sheets may converge on
some degenerate sheet, and if this is the case, then the
coordinate system breaks down at the degenerate sheet.\footnote{It
should be emphasized that only a ``local" radial foliation of an
arbitrarily small spacetime region surrounding $\bar{\cal T}$ is
necessary for the analysis in this work.
The full radial foliation
of $\cal M$ is introduced only to have a closer analogy with temporal
foliations.} With a notation similar to the one introduced above,
we may represent $\bar{\cal T}$ by
$\bar{\cal T}_{r''}$, so the level hypersurface
specified by $r = r''$ is $\bar{\cal T}$ (the inner radial
sheet is $\bar{\cal T}_{r'}$). The spacelike,
outward-pointing, unit, $\bar{\cal T}$
hypersurface normal is denoted
by $\bar{n}$ (the unprimed letter $n$ is reserved for a
related but different vector field introduced below).

{\em Foliations of $\Sigma$ and $\bar{\cal T}$} (see Figure (2)).
It is of interest
to examine how the $\Sigma$ and $\bar{\cal T}$ spacetime foliations
mesh. If a temporal and a radial foliation of ${\cal M}$ are
simultaneously given, then the
intersection
$B_{t,r} \equiv \Sigma_{t} \bigcap \bar{\cal T}_{r}$ is a two
surface with the topology of $B$. Defining
$B_{t} \equiv B_{t,r''}$,
one finds that the family of $B_{t}$ slices foliates
$\bar{\cal T}$. This
foliation of $\bar{\cal T}$ and its generic leaf are
both loosely referred
to as $B$. The timelike, future-pointing, unit, hypersurface normal
of this foliation is $\bar{u}$. In general, the vector fields $u$
and $\bar{u}$ do not
coincide on $\bar{\cal T}$. Fixation of the
time gives a family of sheets
$B_{r} \equiv B_{t_{*}, r}$ which radially foliate the $\Sigma$
hypersurface specified by $t = t_{*}$.
This foliation of $\Sigma$ and its generic leaf
are also represented by $B$. The spacelike, outward-pointing, unit
normal of this foliation is denoted by $n$, and in general $n$ and
$\bar{n}$ do not coincide on $\Sigma$.

{\em Clamped foliations} (see Figure (3)).
Often in this paper we need to consider
a particular subclass of $\Sigma$
foliations, determined by the
property that on $\bar{\cal T}$ the timelike
normal $u$ is orthogonal to $\bar{n}$.
Such foliations are denoted
by $\bar{\Sigma}$ with corresponding normal $\bar{u}$.
(So we have
$\bar{u} \cdot \bar{n} = 0$ on $\bar{\cal T}$, where $\bar{u}$
is also the normal for the $B$ foliation
of $\bar{\cal T}$.) We described a $\bar{\Sigma}$ foliation
as {\em clamped}. Note that
it may not be possible for a temporal foliation to be
clamped over all of $\bar{\cal T}$,
since the $u$ normals of $t'$ and $t''$ may not be orthogonal
to $\bar{n}$ (assuming that $t'$ and $t''$ should
be members of the
family of $\bar{\Sigma}$ slices).
We can also consider the locus of
points which is the Eulerian history of
$B$ with respect to an
(in general) unclamped $\Sigma$ foliation.
This ``boundary", denoted
by ${\cal T}$ is generated by the integral
curves of $u$ and may ``crash
into" or ``emerge from" the actual boundary
$\bar{\cal T}$. 

We maintain this barred and unbarred
notation when it is necessary
to deal with an unclamped $\Sigma$
foliation. However, in $\S$ III.B, which presents the derivation of
the quasilocal densities, we make the {\em clamping assumption}
which means that only clamped $\bar{\Sigma}$ foliations
of spacetime
${\cal M}$ are considered (or every $\Sigma$ foliation
is a $\bar{\Sigma}$
foliation). When the clamping assumption is made, over-bars
become redundant, and therefore {\em in $\S$ III.B we 
drop all bars
from the formalism}. In this subsection of $\S$ III we assume 
that the $u$
normals of $t'$ and $t''$ are orthogonal to the
$\bar{\cal T}$ normal
$\bar{n}$ (in this section denoted simply by
${\cal T}$ and $n$). Though
this is a limiting assumption, it in no way
affects the generality of this paper, as we return to the fully
general scenario in the following section. We demonstrate that
the clamping assumption is a purely kinematical condition.

\subsection{Conventions and notation} We adopt the following index
notation. Lowercase Greek letters serve as ${\cal M}$ spacetime
indices. Lowercase Latin indices from the {\em latter half
of the alphabet} serve as $\Sigma$ (and $\bar{\Sigma}$) indices
and as $\bar{\cal T}$ (and ${\cal T}$) indices.
There is -hopefully- no confusion
caused by this dual use of Latin indices. Lowercase
Latin letters from the {\em first half of the alphabet} serve as
$B$ indices. Orthonormal (or when appropriate pseudo-orthonormal)
labels and indices for each space are represented by the same
letters with hats. For example,
$\hat{\mu}$ is a spacetime tetrad index
and $\hat{a}$ is a $B$ dyad index.

The spacetime metric is $g_{\mu\nu}$ with associated
(metric-compatible and torsion-free) covariant derivative
operator $\nabla_{\mu}$, and $e_{\hat{\mu}}\,^{\sigma}$ denotes
a spacetime tetrad. The (pseudo)orthonormal symbol on spacetime
is defined by
$\epsilon_{\hat{0}\hat{1}\hat{2}\hat{3}} =
1 = - \epsilon^{\hat{0}\hat{1}\hat{2}\hat{3}}$.
Respectively, we have $\bar{\gamma}_{ij}$ and $\bar{\cal D}_{j}$
($\gamma_{ij}$ and ${\cal D}_{j}$), $h_{ij}$ and $D_{j}$
($\bar{h}_{ij}$ and $\bar{D}_{j}$), and
$\sigma_{ab}$ and $\delta_{a}$
denoting the metric and intrinsic
covariant derivative operators on
$\bar{\cal T}$ (${\cal T}$),
$\Sigma$ ($\bar{\Sigma}$), and $B$. We use
$\bar{\xi}_{\hat{r}}\,^{j}$
($\xi_{\hat{r}}\,^{j}$), $E_{\hat{r}}\,^{j}$
($\bar{E}_{\hat{r}}\,^{j}$),
and
$\theta_{\hat{a}}\,^{b}$, respectively, to represent a triad on
$\bar{\cal T}$ (${\cal T}$),
a triad on $\Sigma$ ($\bar{\Sigma}$), and a dyad on $B$.
Respectively,
the permutation symbols on $\bar{\cal T}$
(${\cal T}$), $\Sigma$ ($\bar{\Sigma}$),
and $B$ are defined by
$\bar{\epsilon}_{\hat{0}\hat{1}\hat{2}}
= - 1 = \bar{\epsilon}^{\hat{0}\hat{1}\hat{2}}$
($\epsilon_{\hat{0}\hat{1}\hat{2}} =
- 1 = \epsilon^{\hat{0}\hat{1}\hat{2}}$),
$\epsilon_{\hat{1}\hat{2}\hat{3}} =
1 = \epsilon^{\hat{1}\hat{2}\hat{3}}$
($\bar{\epsilon}_{\hat{1}\hat{2}\hat{3}}
= 1 = \bar{\epsilon}^{\hat{1}\hat{2}\hat{3}}$),
and $\epsilon_{\hat{1}\hat{2}} = 1 =
\epsilon^{\hat{1}\hat{2}}$.
(In Ref.\ \cite{Lau} the convention for the
$\cal T$ orthonormal symbol differs by a sign.)

\subsection{Spacetime decompositions} The foliations just
discussed lead to two decompositions of the spacetime metric.
We choose to examine the metric in a coframe which has
${\rm d}t$ and ${\rm d} r$ as two of
the coframe legs. To begin with, the temporal foliation $\Sigma$
allows the matrix of metric components to be written as
\begin{equation}
\left\| g_{\mu\nu} \right\| = \left(
\begin{array}{cc} - N^{2} + h_{ij}\, V^{i}\,
V^{j} & h_{ij}\, V^{j} \\
h_{ij}\, V^{j} & h_{ij} \end{array} \right)\, ,
\end{equation}
where the $\Sigma$ indices run over $(1,2,{\rm r})$.
The $N$ and $V^{j}$ are
the ordinary ADM lapse and shift.
Further, since each of the $\Sigma$
slices is foliated independently by nested sheets with the
topology of
$B$, the matrix form of the $\Sigma$ three-metric is given by

\begin{equation}
\left\| h_{ij} \right\| =
\left( \begin{array}{cc}
\sigma_{ab} &
\sigma_{ab}\, \beta^{b} \\ \sigma_{ab}\,
\beta^{b} & \alpha^{2} +
\sigma_{ab}\, \beta^{a}\, \beta^{b}
\end{array} \right)\, .
\end{equation}
Here,
$\alpha$ and $\beta^{a}$ are the ``lapse" and ``shift"
associated with the induced radial foliation of $\Sigma$. The
super matrix formed by combining these expressions gives
the so-called $(1+2)+1$ form of the metric. The $(1+2)$
indicates that three-space has been split into a radial
direction plus a two-space, while the $1$ indicates the time
direction.

Similarly, beginning with the full radial foliation
$\bar{\cal T}$ of spacetime one has,

\begin{equation}
\left\| g_{\mu\nu} \right\|
= \left( \begin{array}{cc} \bar{\gamma}_{ij}
& \bar{\gamma}_{ij}\, \bar{\beta}^{j} \\
\bar{\gamma}_{ij}\, \bar{\beta}^{j} &
\bar{\alpha}^{2} + \bar{\gamma}_{ij}\,
\bar{\beta}^{i}\, \bar{\beta}^{j}
\end{array} \right)\, ,
\end{equation}
where the $\bar{\cal T}$ indices run over $({\rm t},1,2)$.
The $\bar{\alpha}$ and $\bar{\beta}^{\alpha}$ are
the gauge variables associated with this foliation. The
submatrix associated with $\bar{\gamma}_{ij}$ is

\begin{equation}
\left\| \bar{\gamma}_{ij} \right\|
= \left( \begin{array}{cc} -
\bar{N}^{2} + \sigma_{ab}\,
\bar{V}^{a}\, \bar{V}^{b} & \sigma_{ab}\,
\bar{V}^{b} \\ \sigma_{ab}\,
\bar{V}^{b} & \sigma_{ab}
\end{array}
\right)\, ,
\end{equation}
where $\bar{N}$ and $\bar{V}^{a}$ are the lapse and
shift associated with the induced $B$
foliation of $\bar{\cal T}$.
The super matrix of components for this splitting is the
metric in $1+(2+1)$ form.

It is a straightforward exercise to express the ``barred"
variables in terms of the ``unbarred" variables by simply
equating the components of the $(1+2)+1$ and $1+(2+1)$
versions of the spacetime metric. First, define

\begin{equation}
v \equiv \frac{V \cdot n}{N} =
\frac{\alpha V^{\rm r}}{N}\,\,\, ;\,\,\, \bar{v} \equiv
\frac{\bar{\beta} \cdot \bar{u}}{\bar{\alpha}} = \frac{\bar{N}
\bar{\beta}^{\rm t}}{\bar{\alpha}}\,\,\,
({\rm where}\,\,\, v = - \bar{v})
\end{equation}
and the point-dependent boost factor
$\gamma = (1 - v^{2})^{-1/2} = (1 - \bar{v}^{2})^{-1/2}$.
With this boost factor the set of transformation
equations may be written as

\begin{eqnarray}
\bar{N} & = & \frac{N}{\gamma} \nonumber \\
\bar{\alpha} & = & \alpha \gamma \nonumber \\
\bar{V}^{b} & = & V^{b} + V^{\rm r}\, \beta^{b}
\label{set} \\
\bar{\beta}^{b} & = & \gamma^{2} \left( \beta^{b} +
\frac{v^{2}}{V^{\rm r}}\, V^{b}\right) \nonumber \\
\bar{\beta}^{\rm t}
& = & - \frac{(v \gamma)^{2}}{V^{\rm r}}\, . \nonumber
\end{eqnarray}
The clamping assumption is tantamount to the
$v \rightarrow 0$
($\gamma \rightarrow 1$) limit on the boundary 
$\bar{\cal T}$, in which case there is no
longer a distinction between barred and unbarred variables.
Note that in this case
$V^{\vdash} \equiv V \cdot n = \alpha\, V^{\rm r} = 0$,
which, as described in \cite{BY}, implies that in the
canonical form of the theory the $\Sigma$ Hamiltonian can not
drive field configurations across the boundary $B$.

\section{Quasilocal stress-energy-momentum densities}

\subsection{Action and variational principle}
Before turning to the derivation of the quasilocal
densities, we must
describe the action principle which is the
cornerstone of our approach. Our starting point is the
first-order {\em Goldberg action}\cite{Goldberg}
 \begin{equation}
S^{\it 1}\left[ e^{\hat{\rho}}\,_{\mu}\right] \equiv
\frac{1}{2\kappa}\int_{\cal M}\Gamma^{\hat{\rho}}\,_{\hat{\tau}}
\wedge e^{\hat{\tau}} \wedge \sigma_{\hat{\rho}}\, , \label{gaction}
\end{equation}
where $\kappa = 8\pi$ (in units with $G = c = 1$) and
$\Gamma_{\hat{\sigma}\hat{\tau} \mu}
= e_{\hat{\sigma} \nu}\,\nabla_{\mu}\, e_{\hat{\tau}}\,^{\nu}$
represent the spacetime
connection one-forms which specify the Levi-Civita connection on
${\cal M}$ with respect to the tetrad
$e^{\hat{\rho}}\,_{\mu}$. Also, the
{\em Sparling two-forms}\cite{Goldberg,Sparling} are defined by

\begin{equation}
\sigma_{\hat{\rho}} \equiv - \frac{1}{2}\,
\epsilon_{\hat{\rho}\hat{\sigma}\hat{\tau}\hat{\mu}}\,
\Gamma^{\hat{\sigma}\hat{\tau}} \wedge e^{\hat{\mu}}\, .
\end{equation}
Therefore, as mentioned, the Goldberg action is not a Palatini
action in  which tetrad $e^{\hat{\rho}}\,_{\mu}$ and connection
$\Gamma^{\hat{\sigma}}\,_{\hat{\tau}\mu}$
are varied independently. As it stands, the
action (\ref{gaction}) possesses superfluous tetrad dependence.
However, note that the Goldberg action is invariant under
spacetime diffeomorphisms
which preserve the boundary, since it is written purely in the
language of differential forms.\cite{DiffGeom}

Our goal is to identify the Goldberg action (\ref{gaction})
with the familiar ``$Tr K$" action \cite{York} used in metric gravity.
The extrinsic curvature tensor associated with the $\Sigma$
foliation is defined by
$K_{\mu\nu} \equiv
- h^{\lambda}_{\nu}\,\nabla_{\lambda}\, u_{\mu}$
(with the projection operator
$h^{\lambda}_{\mu} = g^{\lambda}_{\mu} + u^{\lambda}\,u_{\mu}$),
while the extrinsic
curvature tensor associated with the $\bar{\cal T}$ foliation is
defined by
$\bar{\Theta}_{\mu\nu} \equiv
- \bar{\gamma}^{\lambda}_{\nu}\,\nabla_{\lambda}\, \bar{n}_{\mu}$
(with the projection operator
$\bar{\gamma}^{\lambda}_{\mu} = g^{\lambda}_{\mu}
- \bar{n}^{\lambda}\,\bar{n}_{\mu}$).
The first step towards the desired identification is
to note that the action differs from the ordinary Hilbert
action by a pure divergence \cite{Goldberg,Lau}

\begin{equation}
\frac{1}{2\kappa}\int_{\cal M}\Gamma^{\hat{\rho}}\,_{\hat{\tau}}
\wedge e^{\hat{\tau}}
\wedge \sigma_{\hat{\rho}} =
\frac{1}{2\kappa}\int_{\cal M} \Re e^{*} -
\frac{1}{2\kappa}\int_{\cal M}
{\rm d}\left(e^{\hat{\rho}} \wedge \sigma_{\hat{\rho}}\right)\, ,
\end{equation}
where $e^{*}$ is volume form on ${\cal M}$.
Evidently, all of the action's
tetrad dependence resides exclusively in boundary terms,

\begin{equation}
- \frac{1}{2\kappa}\int_{\cal M} {\rm d}\left(e^{\hat{\rho}}
\wedge \sigma_{\hat{\rho}}\right) = \frac{1}{\kappa}\int_{\cal M}
{\rm d}^{4}x\sqrt{- g}\nabla_{\mu}\left(e^{\hat{\rho} \mu}\,
\nabla_{\lambda}\,e_{\hat{\rho}}\,^{\lambda}\right)\, .
\end{equation}
Now, if the time leg
of the tetrad $e_{\hat{0}}$ coincides with the future-pointing
normals $u$ on both $t'$ and $t''$, then the boundary
terms associated with these manifolds are the desired $TrK$
terms. Likewise, enforcing the condition that the third tetrad
leg $e_{\hat{3}}$ coincides with the $\bar{\cal T}$ normal
$\bar{n}$ on $\bar{\cal T}$ ensures that the one obtains the
desired $Tr\bar{\Theta}$ term for the $\bar{\cal T}$
boundary term. We
assume that the variational set of tetrads obey these
conditions.
However, in general such tetrads are doubled-valued on the
{\em corners}
$B'' \equiv t'' \bigcap \bar{\cal T}$ and
$B' \equiv t' \bigcap \bar{\cal T}$,
since $u \cdot \bar{n}$ need not vanish on these two-surfaces.
Therefore, to express the action (\ref{gaction}) in the desired
form, relax the second gauge condition on
$e_{\hat{3}}$ on a ``small" (not connected)
neighborhood of the corners such that the tetrad is
single-valued. Next, take the limit that this small neighborhood
``shrinks" to just the corners $B'$ and $B''$. Such a limit
procedure is described in {\em Appendix D}, and it yields
the following expression for the action:

\begin{equation}
S^{\it 1} =
\frac{1}{2\kappa}\int_{\cal M}{\rm d}^{4}x\,\sqrt{- g}\,\Re +
\frac{1}{\kappa}\int^{t''}_{t'}{\rm d}^{3}x\,\sqrt{h}\, K\, -
\frac{1}{\kappa}\int_{\bar{\cal T}}
{\rm d}^{3}x\,\sqrt{-\bar{\gamma}}\,
\bar{\Theta} - \frac{1}{\kappa}\int^{B''}_{B'}
{\rm d}^{2}x\,\sqrt{\sigma}\,\phi\, ,\label{cico}
\end{equation}
where $\phi \equiv \tanh^{-1}\, v$ is the point-dependent boost
parameter on $B''$ and $B'$
associated with the boost velocity $v$
defined in the last section. The corner terms were first given
by Hayward and Wong for the metric
action.\cite{Hayward} Heuristically, they arise because, though
the corners constitute a set of measure zero in the $TrK$
integration over all of $\partial{\cal M}$, the trace of the
extrinsic curvature is infinite on
these two-surfaces (as the normal of $\partial{\cal M}$ changes
discontinuously from $u$ to $\bar{n}$). Note that the corner
contributions to the action vanish if the initial and final
slices are clamped to $\bar{\cal T}$. To obtain the variation of
(\ref{cico}), one may straightforwardly vary the action
(\ref{gaction}) and then apply the limiting procedure. This
direct method is sketched in {\em Appendix D}. However, in
the interest of brevity we borrow from results given in
Refs.\ \cite{BYL,Hayward}. Subject to the
chosen ``internal" gauge fixing, the action (\ref{gaction}) is
a tetrad version of the metric action used in Ref.\ \cite{BY}
to define quasilocal stress-energy-momentum in general
relativity. Hence, for the moment we may regard it as a
metric action. Indeed, only the
gauge-invariant quantities $\bar{\gamma}_{ij}$, $h'_{ij}$, and
$h''_{ij}$ are fixed on the boundary $\partial {\cal M}$ in the
associated variational principle. Refs.\ \cite{BYL,Hayward} have
shown that the boundary contributions to the variation of
$S^{\it 1}$ are

\begin{equation}
\left(\delta S^{\it 1}\right)_{\partial{\cal M}}
= \int_{t'}^{t''}{\rm d}^{3}x\,
p^{ij}\,\delta h_{ij}
+ \int_{\bar{\cal T}}{\rm d}^{3}x\,
\bar{\pi}^{ij}\,\delta \bar{\gamma}_{ij}
- \frac{1}{\kappa}\int^{B''}_{B'}{\rm d}^{2}x\, \phi\,
\delta \sqrt{\sigma}\, , \label{parthenon}
\end{equation}
where the gravitational momenta are given by

\begin{eqnarray}
p^{ij} & = & \frac{\sqrt{h}}{2\kappa}
\left(K\, h^{ij} - K^{ij}\right) \nonumber \\
& & \label{momenta} \\
\bar{\pi}^{ij} & = & - \frac{\sqrt{-\bar{\gamma}}}{2\kappa}
\left(\bar{\Theta}\,
\bar{\gamma}^{ij} - \bar{\Theta}^{ij}\right)\, . \nonumber
\end{eqnarray}
The variable $p^{ij}$ becomes the standard ADM momenta
in the canonical form of metric gravity, and it is conjugate
to $h_{ij}$. Likewise,
$\bar{\pi}^{ij}$
is the ADM-type momenta conjugate to $\bar{\gamma}_{ij}$,
but now canonical conjugacy is defined with
respect to $\bar{\cal T}$.
Note that equation (\ref{parthenon}) includes corner
contributions to the variation which feature fixation of
intrinsic geometry, in harmony with the fact that the induced
metric is fixed on $\partial{\cal M}$.

There is a complex-valued action functional, closely related to
(\ref{gaction}), which is based on the {\em self-dual} $(+)$
(or {\em anti-self-dual} $(-)$)
{\em connection forms}

\begin{equation}
\Gamma^{(\pm)\,\! \hat{\rho}\hat{\sigma}} =
\frac{1}{2}\left(\Gamma^{\hat{\rho}\hat{\sigma}} \mp
\frac{i}{2}\epsilon^{\hat{\rho}\hat{\sigma}\hat{\tau}\hat{\mu}}
\Gamma_{\hat{\tau}\hat{\mu}}\right)\, .
\label{winesap}
\end{equation}
This action is referred to as the {\em complex Goldberg action}
and has the form (here we take the self-dual case)

\begin{equation}
S\left[ e^{\hat{\rho}}\,_{\mu}\right]
 = \frac{1}{2\kappa}\int_{\cal M}\Gamma^{\hat{\rho}}\,_{\hat{\tau}}
 \wedge e^{\hat{\tau}} \wedge \sigma^{(+)}\,_{\hat{\rho}}\, ,
 \label{cgaction}
\end{equation}
where the complex Sparling two-forms are

\begin{equation}
\sigma^{(\pm)}\,_{\hat{\rho}} =
- \epsilon_{\hat{\rho}\hat{\sigma}\hat{\tau}\hat{\mu}}\,
\Gamma^{(\pm)\,\!\hat{\sigma}\hat{\tau}} \wedge
e^{\hat{\mu}}\, .
\label{2sd/asdsigSparling}
\end{equation}
The complex action (\ref{cgaction}) differs from the previous
one (\ref{gaction}) by a purely imaginary boundary term.
Indeed, setting

\begin{equation}
S = S^{\it 1} - S^{\it 0}\, , \label{solon}
\end{equation}
we find that

\begin{equation}
- S^{\it 0} =
\frac{1}{2\kappa}\int_{\cal M} {\rm d} \left[ e^{\hat{\rho}}
\wedge \left( \sigma_{\hat{\rho}} -
\sigma^{(+)}\,_{\hat{\rho}}\right)\right]\, .  \label{dahi}
\end{equation}
With the gauge choices made above and the limiting procedure
described in {\em Appendix D}, an
appeal to Stokes' theorem yields

\begin{equation}
- S^{\it 0} =
- \frac{\rm i}{2\kappa}\int^{t''}_{t'}{\rm d}^{3}x\,
\sqrt{h}\,
\epsilon^{\hat{r}\hat{s}\hat{p}}\,\omega_{\hat{s}\hat{p}
j}\, E_{\hat{r}}\,^{j} +
\frac{\rm i}{2\kappa}\int_{\bar{\cal T}}{\rm d}^{3}x\,
\sqrt{-\bar{\gamma}}\,\,
\bar{\epsilon\,}^{\hat{r}\hat{s}\hat{p}}\,
\bar{\tau}_{\hat{s}\hat{p}j}\,
\bar{\xi}_{\hat{r}}\,^{j}\, , \label{phedias}
\end{equation}
where
$\bar{\tau}_{\hat{r}\hat{s} j}
= \bar{\xi}_{\hat{r} k}\,
\bar{\cal D}_{j}\,\bar{\xi}_{\hat{s}}\,^{k}$
and
$\omega_{\hat{r}\hat{s} j} =
E_{\hat{r} k}\, D_{j}\, E_{\hat{s}}\,^{k}$
are respectively the triad connection
coefficients on $\bar{\cal T}$
and $\Sigma$. Notice that $- S^{\it 0}$ contributes no
corner terms to the action and that it serves as a
{\em subtraction term} (a functional of the fixed boundary
data) \cite{BY,BYL} in the broadest
sense (it depends on the boundary data of
$\bar{\cal T}$, $t'$, and $t''$).\footnote{To avoid
confusion, it is
crucial to note that in Refs.
\cite{BY,BYL} the notation $S^{\it 0}$ represents an
arbitrary subtraction term, while in this paper
$S^{\it 0}$
represents the specific term (\ref{phedias}).}
Because of the
triad dependence of the subtraction term,
we do not have the
option of viewing the action
(\ref{cgaction}) as solely a metric action.
Furthermore,
in order to fully remove the superfluous tetrad
dependence
associated with the action
$S$, one would have to completely specify the triad on each
boundary element of $\partial{\cal M}$ (though
we do not chose to
completely do so).

Now consider the boundary-term contributions to
the variation of the action (\ref{cgaction}). Since the plan
is to work with the Ashtekar variables in the canonical form
of the theory, first express the boundary-term contributions
(\ref{parthenon}) to the variation of the action $S^{\it 1}$
in terms of the densitized triads on
$\bar{\cal T}$, $t'$, and $t''$.
(This is easily done with the identity
(\ref{hidentity}) given
below and a similar identity for the
$\bar{\cal T}$ metric and triad.)
Adding this result to the variation of (\ref{phedias}), we
find that

\begin{equation}
\left(\delta S\right)_{\partial{\cal M}} =
{\rm i} \int^{t''}_{t'}
{\rm d}^{3}x\, A^{\hat{r}}\,_{j}\,\,
\delta\left(\sqrt{h}\, E_{\hat{r}}\,^{j}\right)
+ {\rm i} \int_{\bar{\cal T}}\,
{\rm d}^{3} x\,
\bar{{\cal A}}^{\hat{r}}\,_{j}\,\,\delta
\left(\sqrt{-\bar{\gamma}}\,\bar{\xi}_{\hat{r}}\,^{j}\right) -
\frac{1}{\kappa}\int^{B''}_{B'}{\rm d}^{2}x\,\phi\,
\delta \sqrt{\sigma}\, ,  \label{delphi}
\end{equation}
where we have introduced the connections

\begin{eqnarray}
A^{\hat{r}}\,_{j} & = & \frac{1}{\kappa}\left(
\omega^{\hat{r}}\,_{j} - {\rm i}\, K^{\hat{r}}\,_{j}\right)
\equiv \frac{1}{\kappa}\left(
- \frac{1}{2}\,\epsilon^{\hat{r}\hat{s}\hat{p}}\,
\omega_{\hat{s}\hat{p} j} - {\rm i}\, K^{\hat{r}}\,_{j}\right)
\nonumber \\
& &  \label{2sen} \\
\bar{{\cal A}}^{\hat{r}}\,_{j} & = & \frac{1}{\kappa}
\left(\bar{\tau\,}^{\hat{r}}\,_{j}
+ {\rm i}\, \bar{\Theta}^{\hat{r}}\,_{j}\right)
 \equiv \frac{1}{\kappa} \left(
\frac{1}{2}\,\bar{\epsilon\,}^{\hat{r}\hat{s}\hat{p}}\,
\bar{\tau}_{\hat{s}\hat{p} j}
+ {\rm i}\, \bar{\Theta}^{\hat{r}}\,_{j}\right)\, .
\nonumber
\end{eqnarray}
(With these conventions
$\omega^{\hat{r}}\,_{\hat{s} j} =
\epsilon^{\hat{r}}\,_{\hat{p}\hat{s}}\,\omega^{\hat{p}}\,_{j}$
and
$\bar{\tau}^{\hat{r}}\,_{\hat{s} j} =
\bar{\epsilon\,}^{\hat{r}}\,_{\hat{p}\hat{s}}\,
\bar{\tau}^{\hat{p}}\,_{j}$.)
The connection variable $A^{\hat{r}}\,_{j}$
is (up to a factor of $\kappa$) the $\Sigma$ {\em Sen connection},
which becomes the Ashtekar connection in the canonical form
of the theory. Likewise, the second connection
${\cal A}^{\hat{r}}\,_{j}$
is the Sen connection associated with $\bar{\cal T}$. It is a
complexified $SO(2,1)$
connection and enjoy properties completely analogous to the
well-known ones enjoyed by the $\Sigma$ Sen connection. In
particular, in terms of the curvature of
${\cal A}^{\hat{r}}\,_{j}$ one may compactly express the
constraints associated the embedding of
$\bar{\cal T}$ in the Einstein
space ${\cal M}$.\cite{Lau,Lau3} {\em Note that here these
connections are not the canonical Ashtekar connections.} We
have not written down imaginary contributions to the corner
terms which presumably arise from integration by parts on
$\delta S^{\it 0}$ terms. In fact, these vanish, and a
calculation which demonstrates this is
outlined in {\em Appendix D}.

\subsection{Quasilocal densities}
We now present all of the fundamental $B$ tensors which serve as
quasilocal densities describing the
stress-energy-momentum content
of the $\Sigma$ gravitational fields
contained within $B$. We express
these densities in terms of the $\Sigma$
Sen connection and triad. In the next section
when studying the canonical form of the action principle, we
consider the canonical versions of these expressions which are
written in terms of the $\Sigma$ Ashtekar
variables. To begin with, we collect a set
$\{ \varepsilon, j_{a}, s^{ab}\}$
of quasilocal densities which is essentially the same
as that described extensively in the original Ref.\ \cite{BY}. This
set is comprised of an {\em energy} surface density $\varepsilon$,
a {\em tangential momentum} surface density $j_{a}$, and a
{\em spatial stress} surface density $s^{ab}$. We also find the
need to introduce a new set
$\{ j_{\vdash} , \hat{\jmath}_{a} , t^{ab}\}$
of quasilocal densities (also considered in \cite{BYL}), which
is comprised of a {\em normal momentum} surface
density $j_{\vdash}$, a {\em tangential momentum}
surface density $\hat{\jmath}_{a}$
(which turns out to be the same as $j_{a}$), and a
{\em temporal stress} surface density $t^{ab}$. Both sets may be
derived from the gravitational action (\ref{cgaction}) via the
Hamilton-Jacobi method as described in the introduction.
Therefore, we adopt the unifying point of view that {\em any
quasilocal stress-energy-momentum quantity is given by the rate
of change of the classical action} $S_{c\ell}$ {\em corresponding
to some variation in the fixed boundary
data of}
$\partial M = t' \bigcup t'' \bigcup \bar{\cal T}$. However,
we do not explicitly consider the classical action as in
Ref.\ \cite{BY}, since we prefer to ``read off" the geometric
expressions for above densities from the boundary contributions
(\ref{delphi}) to the variation of $S$.

In order to ``read off" the various quasilocal densities from the
boundary terms (\ref{delphi}), we make two assumptions in this
subsection. (i) First, we assume that the $\Sigma$ foliation of
${\cal M}$ is clamped, so that $u \cdot \bar{n} = 0$.
Again, this means that
one may drop all overbars associated with three-boundary
quantities from the formalism. Also, this sets $\phi = 0$ on the
corners. The clamping assumption is made in this section only for
convenience, and we return to the general slicing scenario in the
next section. (ii) Second, we enforce partial gauge fixation of
the triads on the boundary elements of ${\cal M}$. Following Ref.
\cite{Lau}, we require that the
${\cal T}$ triad is time-gauge. This
condition ensures that the ${\cal T}$ piece
$S^{\it 0} |_{\cal T}$ of the subtraction term
{\em is functionally linear in the lapse $N$ and shift $V^{a}$}.
As described in detail in Refs.\ \cite{BY,BYL} this linearity
condition is crucial, because it ensures that the quasilocal
energy density $\varepsilon$ and momentum density
$j_{a}$ depend solely on
the Cauchy data of $\Sigma$. Similarly, the triads on both $t'$
and $t''$ are required to be ``radial-gauge." Essentially this
just requires $E_{\hat{3}}$ to coincide with $n$ at $B$. These
restrictions on the $\bar{\cal T}$,
$t'$, and $t''$ triads ensure that the purely
imaginary piece of the corner contribution to the variation
(\ref{delphi}) vanishes (indeed we have already seen that this
is a condition which follows from how the tetrad has been
selected), and they ensure that the
quasilocal densities to be defined
behave appropriately under boosts.
These points become clear below.
The time-gauge and radial-gauge conditions are defined
and discussed in {\em Appendix A}.
Unlike the clamping assumption (i),
these boundary gauge restrictions (ii) are absolutely
necessary for our formalism. Once we have obtained both the
geometric form and a physical interpretation of each quasilocal
density, we turn in the next section to the issue of how these
densities behave under boosts
and also consider the canonical form
of the action principle when the
$\Sigma$ slicing need not be clamped.

Let us first examine the ${\cal T}$ contribution
to the variation of the
complex Goldberg action with the assumption of a
clamped $\Sigma$
slicing. Subject to the time-gauge condition,
the ${\cal T}$ triad and
cotriad can be expressed (at least locally)
in terms of a $B$ dyad
$\theta_{\hat{a}}\,^{b}$ and codyad $\theta^{\hat{a}}\,_{b}$,

\begin{eqnarray}
\xi_{\bot} = 1/N\left(\partial /\partial
t - V^{a}\, \partial /\partial x^{a}\right) & \hspace{1cm} &
\xi_{\hat a} = \theta_{\hat{a}}\,^{b}\, \partial /\partial x^{b}
\nonumber \\
& & \label{bdtriad}\\
\xi^{\bot} = N\, {\rm d}t\hspace{2.55cm}
& \hspace{1cm} & \xi^{\hat a} =
\theta^{\hat{a}}\,_{b}
\left({\rm d}x^{b} + V^{b}\, {\rm d}t\right)\, .
\nonumber
\end{eqnarray}
The time-gauge condition has been indicated by replacing the
triad label $\hat{0}$ with $\bot$. The associated time-gauge
${\cal T}$ connection coefficients are the following:

\begin{eqnarray}
\tau_{\hat{c}\bot\bot} & = & \theta_{\hat{c}}\left[\log N\right]
\nonumber \\
& & \nonumber \\
\tau_{\bot\hat{a}\hat{c}} & = & 1/N\left[
\sigma_{bd}\,\theta_{(\hat{a}}\,^{b}\,
\dot{\theta}_{\hat{c})}\,^{d} +\,
\theta_{\hat{c}}\,^{b}\,\theta_{\hat{a}}\,^{d}\,
\delta_{(b}\, V_{d)}
\right]
\nonumber \\
& &    \label{2papaya} \\
\tau_{\hat{a}\hat{c}\hat{b}} & = &
  \theta_{\hat{b}}\,^{c}\,\theta_{\hat{a}\, b}\, \left(
\delta_{c}\,\theta_{\hat{c}}\,^{b}\right) \nonumber \\
& & \nonumber \\
\tau_{\hat{1}\hat{2}\bot} & = &
1/N\left[\sigma_{bd}\,\theta_{[\hat{1}}\,^{b}\,
\dot{\theta}_{\hat{2}]}\,^{d}
- 1/2\,\epsilon^{ac}\,\delta_{a} V_{c} -
V^{b}\,\tau_{\hat{1}\hat{2} b}\right]\, , \nonumber
\end{eqnarray}
where for this set the ``dot" represents partial time
differentiation.
Plugging these coefficients into
$\left. S^{\it 0}\right|_{\cal T}$, one can verify that
$\left. S^{\it 0}\right|_{\cal T}$ is functionally linear
in the shift $V^{a}$ and has no $N$ dependence. Next, applying
the identities [$\eta_{\hat{r}\hat{s}} = \eta^{\hat{r}\hat{s}}
= diag (-1,1,1)$]
\begin{eqnarray}
\partial\widetilde{\xi}_{\hat{r}}\,^{j}/\partial N & = &
\sqrt{\sigma}\,\theta_{\hat{a}}\,^{j}\,\eta^{\hat{a}}_{\hat{r}}
\nonumber \\
 & & \nonumber \\
\partial\widetilde{\xi}_{\hat{r}}\,^{j}/\partial V^{b} & = &
-\,\sqrt{\sigma}\,\eta^{\bot}_{\hat{r}}\,\sigma^{j}_{b}
\label{triadsplit} \\
& & \nonumber \\
\partial\widetilde{\xi}_{\hat{r}}\,^{j}/
\partial \theta^{\hat{a}}\,_{b}
& = & N\,\sqrt{\sigma}\left( \xi_{\hat{r}}\,^{j}\,
\theta_{\hat{a}}\,^{b} - \eta^{\hat{c}}_{\hat{r}}\,
\theta_{\hat{c}}\,^{b}\,\theta_{\hat{a}}\,^{j}\right)\, .
\nonumber
\end{eqnarray}
to the ${\cal T}$ piece of the boundary variation (\ref{delphi}),
we write the ${\cal T}$ contribution as

\begin{equation}
\left( \delta S \right)_{\cal T} =
- \int_{\cal T} {\rm d}^{3}x\,
\sqrt{\sigma}\left[ \varepsilon\, \delta N -
{j}_{a}\, \delta V^{a} - \frac{N}{2}\,
s^{ab}\, (\delta\theta)_{ab}\right]\, .
\label{ajax}
\end{equation}
Here the quasilocal density $s^{ab}$ is defined with respect to

\begin{equation}
s_{\hat{c}}\,^{b} \equiv \left. \frac{1}{\sqrt{- \gamma}}\,
\frac{\delta S}{\delta\theta^{\hat{c}}\,_{b}}\right|_{\cal T}\, ,
\label{1juniper}
\end{equation}
via
$s^{ab} = s_{\hat{c}}\,^{b}\, \theta^{\hat{c} a}$,
and the expression
$(\delta\theta)_{ab}$ is shorthand for
$2\,\theta_{\hat{a} a}\,\delta\theta^{\hat{a}}\,_{b}$.
Notice that
$(\delta\theta)_{(ab)} = \delta \sigma_{ab}$,
while $(\delta\theta)_{[ab]}$ is a pure gauge variation of the
$B$ dyad. Also, note that $s^{[ab]}$ is
completely determined by the subtraction term
$\delta S^{\it 0} |_{\cal T}$
($s^{(ab)}$ is determined by
$\delta S^{\it 1}$ and $\delta S^{\it 0}$
contributions). Explicitly we have

\begin{eqnarray}
\varepsilon & \equiv & - \left.\frac{1}{\sqrt{\sigma}}\,
\frac{\delta S}{\delta N}\right|_{\cal T} =
- {\rm i}\,{\cal A}^{\hat{a}}\,_{b}\,
\theta_{\hat{a}}\,^{b}\nonumber\\
& & \nonumber \\
{j}_{a} & \equiv & \left.\frac{1}{\sqrt{\sigma}}\,
\frac{\delta S}{\delta V^{a}}\right|_{\cal T} =
- {\rm i}\, {\cal A}^{\bot}\,_{a}
\label{2historyset} \\
& &  \nonumber \\
s^{ab} & \equiv &
\left.\frac{1}{\sqrt{-\gamma}}\,\theta^{\hat{c} a}\,
\frac{\delta S}{\delta \theta^{\hat{c}}\,_{b}}\right|_{\cal T}
= {\rm i}\left({\cal A}^{\hat{r}}\,_{j}\,
\xi_{\hat{r}}\,^{j}\,\sigma^{ab}
- {\cal A}^{\hat{c}}\,_{d}\,\sigma^{d a}\,
\theta_{\hat{c}}\,^{b}\right)\, . \nonumber
\end{eqnarray}
We can rewrite these densities in terms of the $\Sigma$
Sen connection.
The appendix
results (\ref{clampedsen}) express the
time-gauge ${\cal T}$ Sen connection
${\cal A}^{\hat{r}}\,_{j}$ in terms
of the radial-gauge $\Sigma$
Sen connection $A^{\hat{r}}\,_{j}$ and
other gauge variables. Insertion of the appendix results
into the above expressions gives the following:

\begin{eqnarray}
\varepsilon & \equiv &
\epsilon^{\hat{a}\hat{c}}\,
A_{\hat{a} b}\,\theta_{\hat{c}}\,^{b}\nonumber\\
& & \nonumber \\
j_{a} & \equiv &
{\rm i}\, A^{\vdash}\,_{a} \label{3historyset} \\
& &  \nonumber \\
s^{ab} & \equiv &  \epsilon^{\hat{a}\hat{c}}\,
A_{\hat{a} d}\,\sigma^{da}\,
\theta_{\hat{c}}\,^{b} + \left(2{\rm i}/\kappa\,
\Gamma^{(+)}\,_{\hat{1}\hat{2}\bot}
- \epsilon^{\hat{a}\hat{c}}\,
A_{\hat{a} d}\,
\theta_{\hat{c}}\,^{d} \right)\sigma^{ab}\, .
\nonumber
\end{eqnarray}
{\em Henceforth, we assume that $\varepsilon$, $j_{a}$, and
$s^{ab}$ represent these expressions.}
Notice that $\varepsilon$ and $j_{a}$ are built exclusively from
$\Sigma$ Cauchy data
$(E_{\hat{r}}\,^{j}\, ,\, K^{\hat{r}}\,_{j})$.
Because of this fact, $\varepsilon$ and $j_{a}$
can be interpreted as canonical expressions depending on the
Ashtekar variables. Because of the presence of
$2 {\rm i}\, \Gamma^{(+)}\,_{\hat{1}\hat{2}\bot} =
a_{j}\, n^{j} + {\rm i}\,\tau_{\hat{1}\hat{2}\bot}$,
the density $s^{ab}$ does not depend solely on $\Sigma$ Cauchy
data. This term contains the spacetime acceleration
$a^{\mu} = u^{\nu}\,\nabla_{\nu}\, u^{\mu}$
of $u^{\mu}$ as well as the ${\cal T}$ connection coefficient
$\tau_{\hat{1}\hat{2}\bot}$, which describes
the rotation of the $B$ dyad under parallel transport along
the integral curves of $u$. Both of these terms depend on
how the Cauchy data evolve in time. The real parts of the
densities in the above set correspond exactly to the quasilocal
densities first introduced in Ref.\ \cite{BY}. Indeed, expressed
in full detail,

\begin{eqnarray}
\varepsilon & = & \frac{1}{\kappa}\, k \nonumber\\
& & \nonumber \\
j_{a} & = &
- \frac{2}{\sqrt{h}}\, n_{i}\,\sigma_{aj}\,  p^{ij}
- \frac{{\rm i}}{\kappa}\, \omega_{\hat{1}\hat{2} a}
\label{Tset} \\
& &  \nonumber \\
s^{ab} & = & \frac{1}{\kappa}\left[ k^{ab}
+ \left(a_{j}\, n^{j} - k \right) \sigma^{ab}\right]
+ \frac{\rm i}{\kappa}\left( l_{c}\,^{a}\, \epsilon^{bc}
+ \tau_{\hat{1}\hat{2}\bot}\,\sigma^{ab}\right)\, ,
\nonumber
\end{eqnarray}
where $k_{ab}$ is the extrinsic curvature of $B$ as embedded
in $\Sigma$ (with $k = \sigma^{ab}\, k_{ab}$) and $l_{ab}$ is
the extrinsic curvature of $B$ as embedded in ${\cal T}$.

We assume that each density has the same physical
interpretation as given in Ref.\ \cite{BY} and review these
interpretations now. (For the following interpretations to be
valid, one should consider the densities
$\varepsilon$, $j_{a}$, and
$s^{ab}$ to be evaluated ``on-shell",
i.\ e.\ evaluated on some particular solution of the Einstein
field equations.) From its definition
$\sqrt{\sigma}\,\varepsilon$
equals minus the time rate of change of the action $S$, where
the time separation between the $B$ slices of $\cal T$ is
controlled by the lapse $N$ on $\cal T$ (fixed as boundary
data in the variational principle). Therefore, $\varepsilon$ is
interpreted as an energy surface density for the system
as measured by the Eulerian observers of $\Sigma$ at $B$.
The total {\em quasilocal energy}
associated with the $\Sigma$ gravitational fields is

\begin{equation}
E = \int_{B}{\rm d}^{2}x\,\sqrt{\sigma}\,\varepsilon\, ,
\end{equation}
the integral of the quasilocal energy density over the
two-surface $B$.
This energy is the value of the on-shell
Hamiltonian\footnote{Whether or not it is
possible to find a truly satisfactory Hamiltonian for a
spatially bounded slice $\Sigma$ is a subtle issue in its own
right. Following Ref.\ \cite{BY}, this paper assumes that the
correct Hamiltonian for a bounded region is the one which is
``read off" from the canonical form of the gravitational
action appropriate for a spatially bounded spacetime region.}
which corresponds to the choice $N = 1$ and $V^{k} = 0$ on $B$,
and it is a functional on the gravitational phase space associated 
with $\Sigma$. In a similar fashion, $j_{a}$ is interpreted as a
tangential-momentum surface density, and the integral

\begin{equation}
J_{\phi} = \int_{B}{\rm d}^{2}x\,\sqrt{\sigma}\,\phi^{a}\,j_{a}
\end{equation}
is a total tangential-momentum quantity associated with the
$\Sigma$ gravitational fields and the $B$ vector field $\phi^{a}$. 
When $\phi^{a}$ is a rotational Killing field whose orbits are 
contained in $B$, 
the real part of $J_{\phi}$ rigorously represents the total 
quasilocal $\Sigma$ angular momentum.\cite{BY} 
On-shell, the integral $J_{\phi}$ is minus the value of the Hamiltonian
which corresponds to the choice $N = 0$, $V^{\vdash} = 0$, and
$V^{k}\sigma_{k}^{a} = \phi^{a}$ on the boundary. 
In the special case when $\phi^{a}$ is a genuine rotational 
Killing field, the form of $j_{a}$ makes it tempting 
to identify the imaginary part of $J_{\phi}$ with the ``spin" of 
the $B$ dyad.\cite{Yorksuggestion} 
Finally, note that the real part of $s^{ab}$ represents the flux 
of the $a$ component of momentum in the $b$ direction.\cite{BY}

The original set of quasilocal densities have been obtained
from a careful analysis of the ${\cal T}$ contribution to the
variation (\ref{delphi}) of the action. In a similar
fashion we now analyze the $t'$ and $t''$ contributions
to the variation.\footnote{The
remainder of this section is based on Refs.
\cite{BYL,JDBnotes}.} Often we drop the $'$ and $''$ notations
with the understanding that all expressions may refer to
either the manifold $t'$ or $t''$. Remember that the $t'$ and
$t''$ triads are radial-gauge. The radial
gauge is indicated by replacing the triad label $\hat{3}$ by
$\vdash$. Therefore, with may split
$\widetilde{E}_{\hat{r}}\,^{j}$ into $\alpha$ and $\beta^{a}$
(the gauge variables associated with the $1+2$ split of
$h_{ij}$) as well as the $B$ codyad $\theta^{\hat{a}}\,_{b}$.
With this assumption, we find identities like
those in (\ref{triadsplit}). Therefore, it easy to write the
$t'$ and $t''$ contributions to the
variation (\ref{delphi}) as

\begin{equation}
\left( \delta S\right)^{t''}_{t'} =
- \int^{t''}_{t'}{\rm d}^{3}x\,
\sqrt{\sigma}\left[ j_{\vdash}\,\delta \alpha
+ \hat{\jmath}_{a}\, \delta \beta^{a} -
\frac{\alpha}{2}\, t^{ab}\, (\delta\theta)_{ab}\right]\, .
\end{equation}
The new quasilocal densities described at the beginning of
this subsection are then

\begin{eqnarray}
j_{\vdash} & \equiv & - \left.\frac{1}{\sqrt{\sigma}}\,
\frac{\delta S}{\delta \alpha}\right|_{t''} = - {\rm i}\,
A^{\hat{a}}\,_{b}\,\theta_{\hat{a}}\,^{b} \nonumber \\
& & \nonumber \\
\hat{\jmath}_{a} & \equiv & - \left.\frac{1}{\sqrt{\sigma}}\,
\frac{\delta S}{\delta \beta^{a}}\right|_{t''} = {\rm i}\,
A^{\vdash}\,_{a} \label{capset}  \\
      & & \nonumber \\
t^{ab} & \equiv & \left.\frac{1}{\sqrt{h}}\,
\theta^{\hat{c} a}\,
\frac{\delta S}{\delta \theta^{\hat{c}}\,_{b}}
\right|_{t''} = {\rm i}
\left(A^{\hat{r}}\,_{j}\, E_{\hat{r}}\,^{j}\,\sigma^{ab} -
A^{\hat{a}}\,_{c}\,\sigma^{ca}\,
\theta_{\hat{a}}\,^{b}\right)\, ,
\nonumber
\end{eqnarray}
with the same expressions for the
densities associated with the
manifold $t'$. In full detail these are

\begin{eqnarray}
j_{\vdash} & = &
- \frac{2}{\sqrt{h}}\, n_{i}\, n_{j}\, p^{ij}
\nonumber \\
& & \nonumber \\
\hat{\jmath}_{a} & = &
- \frac{2}{\sqrt{h}}\, n_{i}\, \sigma_{aj}\, p^{ij}
- \frac{\rm i}{\kappa}\,\omega_{\hat{1}\hat{2}a}
\label{Sset}\\
      & & \nonumber \\
t^{ab} & = &
\frac{2}{\sqrt{h}}\,
\sigma^{a}_{i}\,\sigma^{b}_{j}\, p^{ij}
- \frac{\rm i}{\kappa}
\left(\omega_{\hat{1}\hat{2}\vdash}\,\sigma^{ab}
+ k_{c}\,^{a}\, \epsilon^{bc} \right)\, .
\nonumber
\end{eqnarray}
One could also write $j_{\vdash} = - (1 /\kappa) l$,
where $l$ is the trace of $l_{ab}$.
Note that the definitions
of $j_{a}$ and $\hat{\jmath}_{a}$ are identical, and hence
$\hat{\jmath}_{a}$ carries the same physical interpretation
as $j_{a}$. Therefore, from now on we suppress the hat on
$\hat{\jmath}_{a}$. This equivalence of
$\hat{\jmath}_{a}$ with $j_{a}$ results from the
chosen gauge conditions. Also a result of these conditions
is the fact that both $j_{\vdash}$ and $\varepsilon$
are real. It
turns out that the reality of
$j_{\vdash}$ and $\varepsilon$ (or equivalently
that the subtraction
term $S^{\it 0}$ has no $\alpha$ or $N$ dependence) is quite
crucial, as it ensures that $j_{\vdash}$ and
$\varepsilon$ behave
well under boosts. Note that even if we had
not enforced
the radial-gauge condition on the $t'$ and
$t''$ triads,
then all of the densities listed immediately
above would
still by construction depend only on $\Sigma$
Cauchy data. As
shown in \cite{BYL}, $j_{\vdash}$ is a normal momentum
density, and the total normal momentum associated with
the $\Sigma$ fields is given by

\begin{equation}
J_{\vdash}
= \int_{B}{\rm d}^{2}x\,\sqrt{\sigma}\,j_{\vdash}\, .
\end{equation}
This expression is minus the value of the on-shell Hamiltonian
which corresponds to the choice $N = 0$, $V^{k}\sigma_{k}^{a} = 0$, and
$V^{\vdash} = \alpha\, V^{\rm r} = 1$ on $B$ (heuristically,
we may think of $J_{\vdash}$ as an on-shell value
of the Hamiltonian which generates unit dilations of the
system). Finally, we refer to $t^{ab}$ as the temporal
stress density but lack a precise physical interpretation for
this density.

\section{Boosted densities and the canonical action}

\subsection{Boost relations and invariants}
We now demonstrate that our collection of quasilocal
densities behave
under generalized boosts  in a manner which is in
accord with the
equivalence principle. Fix a spacelike two-surface
$B$ in spacetime
and also consider an arbitrary spacelike hypersurface
$\bar{\Sigma}$
which has boundary $\partial \bar{\Sigma} = B$.
The hypersurface normal of $\bar{\Sigma}$ is $\bar{u}$.
If we view the $\bar{\Sigma}$ slice as a member of a
temporal foliation, then we
may define the Eulerian history of $B$ as $\bar{\cal T}$.
By construction $\bar{\Sigma}$ is clamped to $\bar{\cal T}$.
The observers at $B$ who are instantaneously at rest in the
$\bar{\Sigma}$ slice (Eulerian observers of
$\bar{\Sigma}$) determine
the following set of quasilocal densities:
\begin{eqnarray}
\bar{\varepsilon} & = & \epsilon^{\hat{a}\hat{c}}\,
\bar{A}_{\hat{a} b}\,\theta_{\hat{c}}\,^{b}\nonumber\\
& & \nonumber \\
\bar{\jmath}_{\vdash'} & = & - {\rm i}\,
 \bar{A}^{\hat{a}}\,_{b}\,\theta_{\hat{a}}\,^{b} \nonumber \\
& &  \nonumber \\
\bar{\jmath}_{a} & = & {\rm i}\, \bar{A}^{\vdash'}\,_{a}
\label{barreddensities} \\
 & & \nonumber \\
\bar{s}^{ab} & = & \epsilon^{\hat{a}\hat{c}}\,
\bar{A}_{\hat{a} d}\,\sigma^{da}\,
\theta_{\hat{c}}\,^{b} + \left(2 {\rm i}/\kappa\,
\Gamma^{(+)}\,_{\hat{1}\hat{2}\bot'} -
\epsilon^{\hat{a}\hat{c}}\, \bar{A}_{\hat{a} d}\,
\theta_{\hat{c}}\,^{d} \right)\sigma^{ab}
\nonumber \\
      & & \nonumber \\
\bar{t}^{ab} & = &  {\rm i}
\left(\bar{A}^{\hat{r}}\,_{j}\,
\bar{E}_{\hat{r}}\,^{j}\,\sigma^{ab}
- \bar{A}^{\hat{a}}\,_{c}\,
\sigma^{ca}\,\theta_{\hat{a}}\,^{b}\right)\, .
\nonumber
\end{eqnarray}
The primes appear on some labels in these formulae
because in the triad formalism we
set $\bar{\xi}_{\bot'} = \bar{u}$ and
$\bar{E}_{\vdash'} = \bar{n}$.
\footnote{At this point, the notation
seems overly cluttered, but
its use pays off in the appendix where
the use of marked indices
streamlines some derivations. In our
notation {\em spacetime}
quantities like
$\Gamma^{(+)}\,_{\hat{1}\hat{2}\bot'}$
are not barred. As the prime indicates,
the $\Gamma^{(+)}\,_{\hat{1}\hat{2}\bot'}$
in (\ref{barreddensities}) and the
$\Gamma^{(+)}\,_{\hat{1}\hat{2}\bot}$
in (\ref{3historyset}) are associated with two {\em different}
spacetime tetrads. The labels $\hat{1}$ and $\hat{2}$
on the $\Gamma^{(+)}\,_{\hat{1}\hat{2}\bot'}$ need not
carry primes as the unprimed tetrad and
primed tetrad share these
two $B$ legs, i.\ e.\
$e_{\hat{1}'} = e_{\hat{1}}$ and $e_{\hat{2}'}
= e_{\hat{2}}$ though in general $e_{\bot'} \neq e_{\bot}$.
See the first two paragraphs of {\em Appendix B} for
a fuller explanation of the notation.}
Note that here
$2{\rm i}\,\Gamma_{\hat{1}\hat{2} \bot'}
= \bar{a}_{j}\, \bar{n}^{j} + {\rm i}\,
\bar{\tau}_{\hat{1}\hat{2}\bot'}$.
Now consider a different hypersurface $\Sigma$ which spans $B$
(so like before $\partial\Sigma = B$). We may view
$\Sigma$ as a particular leaf of a temporal
foliation which is not
clamped to $\bar{\cal T}$, the Eulerian
history of $B$ with respect to
$\bar{\Sigma}$. Geometrically, the scenario now is
identical to
the bounded spacetime region $\cal M$ that we have
considered
in the preliminary section. The observers at $B$
who are at
rest in the $\Sigma$ hypersurface determine the
set of quasilocal
densities which are listed in (\ref{3historyset})
and (\ref{capset}) (simply the ``unbarred" versions of
the expressions above). We seek the transformation rules between
the ``barred" and ``unbarred" densities, or, in other words,
the behavior of the quasilocal expressions under
switches of the hypersurface spanning $B$. With the appendix
results (\ref{sensplit}) and (\ref{sdsplit}), it is quite a
simple matter to establish that
\begin{eqnarray}
\bar{\varepsilon} & = & \gamma\,\varepsilon
- v\,\gamma\, j_{\vdash} \nonumber \\
& & \nonumber \\
\bar{\jmath}_{\vdash'} & = &
\gamma\, j_{\vdash} - v\,\gamma\,
\varepsilon \nonumber \\
& & \nonumber \\
\bar{\jmath}_{a} & = & j_{a}
- \frac{\gamma^{2}}{\kappa}\, \delta_{a} v
\label{boostlaws} \\
& & \nonumber \\
\bar{s}^{ab} & = & \gamma\, s^{ab} - v\,\gamma\, t^{ab}
+ \frac{1}{\kappa}\,\sigma^{ab}\,\gamma^{3}\, u[v] +
\frac{1}{\kappa}\,\sigma^{ab}\,v\,\gamma^{3}\, n[v]
\nonumber \\
& & \nonumber \\
\bar{t}^{ab} & = & \gamma\, t^{ab} - v\,\gamma\, s^{ab}
- \frac{1}{\kappa}\,\sigma^{ab}\,\gamma^{3}\, n[v] -
\frac{1}{\kappa}\,\sigma^{ab}\,v\,\gamma^{3}\, u[v]\, .
\nonumber
\end{eqnarray}
These are precisely the Eulerian-Eulerian
boost relations found in Ref.\ \cite{BYL}.
Remarkably, the particular form of the subtraction term
(\ref{dahi}), subject to the chosen gauge fixation, does not
affect the boost relations of the ``bare" densities. It must be
stressed that if the gauge conditions (time gauge on $\cal T$,
and radial gauge on $t'$ and $t''$) had not been enforced when
defining the set of quasilocal densities, then the above boost
relations would not have held. In particular,
if $\varepsilon$ and
$j_{\vdash}$ are defined with subtraction-term contributions
(which in this paper means they would no longer be real), then
the first two boost relations are modified. For an 
application of the first boost relation to the Schwarzschild
geometry see Ref.\ \cite{Lau4}.

Following Refs.\ \cite{JDBnotes} and \cite{BYL},
define trace-free
parts of $s^{ab}$ and $t^{ab}$,
\begin{eqnarray}
\eta^{ab} & \equiv &
s^{ab} - 1/2\, s^{c}\,_{c}\,\sigma^{ab}
= \frac{1}{\kappa}\left(k^{ab} - \frac{1}{2}\,
k\,\sigma^{ab}
+ {\rm i}\, l_{c}\,^{a}\,\epsilon^{bc}\right)
\nonumber \\
& & \\
\zeta^{ab} & \equiv &
t^{ab} - 1/2\, t^{c}\,_{c}\,\sigma^{ab}
= - \frac{1}{\kappa}\left( l^{ab}
-  \frac{1}{2}\, l\,\sigma^{ab}
+ {\rm i}\, k_{c}\,^{a}\,\epsilon^{bc}\right)
\nonumber \, ,
\end{eqnarray}
Notice that the {\em shear stress} $\eta^{ab}$ depends only on
$\Sigma$ Cauchy data. Of course,
$\theta \equiv s^{a}\,_{a}
= 1/\kappa\, (2\, a_{\mu}\, n^{\mu} - k + 2 {\rm i}\,
\tau_{\hat{1}\hat{2}\bot})$
depends on how $u^{\mu}$ and the $B$ dyad are
extended into the future. Similarly, though $t^{ab}$ depends
only on $\Sigma$ Cauchy data, its trace
$\vartheta \equiv t^{a}\,_{a}
= 1/\kappa\, (2\, b_{\mu}\, u^{\mu} + l - 2 {\rm i}\,
\omega_{\hat{1}\hat{2}\vdash})$
depends on how $n$ and the $B$ dyad are extended into the
interior of $B$. We refer to $\zeta^{ab}$ as the
{\em shear temporal stress}. Each of the densities
$\varepsilon$, $j_{\vdash}$, $j_{a}$, $\eta^{ab}$
and $\zeta^{ab}$
depend only on the extrinsic and intrinsic
geometry of $B$ and
the normal $n^{\mu}$ {\em at} $B$.\footnote{I thank
J.\ D.\ Brown
for making this point.} One can easily see that
\begin{eqnarray}
\bar{\eta}^{ab} & = & \gamma\,\eta^{ab}
- v\,\gamma\, \zeta^{ab}
\nonumber \\
& &  \\
\bar{\zeta}^{ab} & = & \gamma\, \zeta^{ab}
- v\,\gamma\, \eta^{ab}\, .
\nonumber
\end{eqnarray}
With the set
$\{\varepsilon , j_{\vdash} , j_{a} , \eta^{ab} ,
\zeta^{ab}\}$
of quasilocal densities we can construct several invariants.
For instance, notice that under boosts the density $j_{a}$
transforms like a gauge potential, since
$\gamma^{2}\,\delta_{a} v = \delta_{a} \phi$.
Therefore, the ``field strength" or curvature
$F_{ab} = 2\,\delta_{[a}\, j_{b]}$ of $j_{b}$
is an invariant.\cite{Szabados} Borrowing
the results from \cite{BYL,JDBnotes}, we write down the
following list of simple invariants
\begin{eqnarray}
^{*}F & = &   \epsilon^{ab} \delta_{a} j_{b} \nonumber  \\
m^{2} & = & \varepsilon^{2} - j_{\vdash}^{\, 2}
\nonumber \\
(m_{\it 1})^{2} & = &
\eta_{ab}\,\eta^{ab} - \zeta_{ab}\,\zeta^{ab}   \\
(m_{\it 2})^{2} & = &
2 {\rm i}\, \epsilon^{ab}\, \eta^{c}\,_{a}\,\zeta_{bc}
= 1/2\, m^{2} - (m_{\it 1})^{2}\, .
\nonumber
\end{eqnarray}
We make no claim that any of these invariants need be
positive. One can also
construct quartic-type invariants. In the metric
formalism where all
of the quasilocal densities are real,
$m^{2}$, $(m_{\it 1})^{2}$, and $(m_{\it 2})^{2}$
are linearly independent. For our theory, the real invariant
$m^{2}$ along with the real and imaginary pieces of
$(m_{\it 1})^{2}$ comprise a linearly independent set.
It is natural to add to this list the scalar curvature
$\cal R$ of $B$. One invariant of interest
which we can build is
\begin{equation}
\sigma^{\mu\sigma}\,\sigma^{\lambda\kappa}\,
C_{\mu\lambda\sigma\kappa} = k^{ab}\, k_{ab} - k^{2}
- l^{ab}\, l_{ab} + l^{2} + {\cal R} =
- \kappa^{2}/2\left[m^{2} + Re(m_{\it 2})^{2}
- 2\kappa^{-2}\, {\cal R}\right]\, ,
\end{equation}
where $C_{\mu\lambda\sigma\kappa}$ is the Weyl tensor of
$g_{\mu\nu}$ (here
we are assuming vacuum), the two-metric
$\sigma_{\mu\nu} = g_{\mu\nu} - n_{\mu}\, n_{\nu}
+ u_{\mu}\, u_{\nu}$ here serves as
the projection operator into the $B$ slices,
and $Re$ stands for ``real part."
This equation is a geometric identity associated with
the embedding
of the two-surface $B$ in spacetime
$\cal M$.\cite{Szabados,S.Hayward,Spivak}
As described in Ref.\ \cite{S.Hayward},
the Bondi mass arises as the appropriate null
asymptotic limit of the quantity
\begin{equation}
M = \frac{1}{\kappa}\,\sqrt{\frac{A}{16\pi}}\int_{B}
{\rm d}^{2}x\,\sqrt{\sigma}\,
\sigma^{\mu\sigma}\,\sigma^{\lambda\kappa}\,
C_{\mu\lambda\sigma\kappa}\, .
\end{equation}
The factor $A$ is the area of $B$, and it has been inserted
in order that the above
expression has units of energy. Moreover, since the invariant 
$m^{2} = (8/\kappa^{2}) \mu\rho$ is essentially the product 
of the convergences $\rho$ and $-\mu$ of the null normals
$(u \pm n)/\sqrt{2}$ to $B$, another invariant we can
built is the Hawking mass\cite{Hawking,S.Hayward}
\begin{equation}
M_{H} = \frac{1}{\kappa}\,\sqrt{\frac{A}{16\pi}}\int_{B}
{\rm d}^{2}x\sqrt{\sigma}
\left[ - (\kappa^{2}/2) m^{2} + {\cal R}\right]\, .
\end{equation}
The Hawking mass also yields the Bondi mass in a suitable
limit.\cite{S.Hayward}

\subsection{Canonical action}
Our goal in this subsection is to consider the variational
principle associated with the canonical form of the action
$S$ (\ref{solon}). In order to express
$S$ in canonical form, we first consider the (3+1) form of
$S$. Begin by expressing $S^{\it 1}$ in (3+1) form. This
can be done straightforwardly with standard methods. However, 
a faster way to get the (3+1) form is to borrow the results 
from Ref.\ \cite{BYL,Hayward,Lau3}. In those
references the action $S^{\it 1}$, viewed as a metric action,
has been expressed in canonical form.\footnote{For the
metric action the (3+1) form is obtained from the canonical
form by simply assuming that $p^{ij}$ has the form given
in (\ref{momenta}) and that
$2 N K_{ij} = - \dot{h}_{ij} + D_{i}\, V_{j}
+ D_{j}\, V_{i}$.}
Therefore, we may simply cast this result into the
language of triads. This is achieved by
assuming that the $\Sigma$ metric is a secondary quantity
derived from $\widetilde{E}_{\hat{r}}\,^{j}$ and by using the
identity
\begin{equation}
\partial h_{ij} /\partial \widetilde{E}_{\hat{r}}\,^{k}
= (E)^{-1} (h_{ij}\, E^{\hat{r}}\,_{k}
- h_{kj}\, E^{\hat{r}}\,_{i}
- h_{ik}\, E^{\hat{r}}\,_{j})\, .
\label{hidentity}
\end{equation}
In anycase, the result is
\begin{equation}
S^{\it 1} = \int_{\cal M}{\rm d}^{4}x
\left[\frac{1}{\kappa}\,
K^{\hat{r}}\,_{j}\,\dot{\widetilde{E}\,}_{\hat{r}}\,^{j}
- N\,{\cal H} - V^{j}\, {\cal H}_{j}\right]
+ \int_{\bar{\cal T}}{\rm d}^{3}x
\left[- \frac{\phi}{\kappa}\,
\dot{\sqrt{\sigma}\,\,\,\,\,}\!\!\!\!\! -
\bar{N}\, \bar{\cal H}^{\it 1} - \bar{V}^{b}\,
\bar{\cal H}^{\it 1}_{b}\right]\, , \label{pylos}
\end{equation}
where we have the following:
\begin{eqnarray}
{\cal H} & = & \frac{1}{2\kappa}\left[ h^{-1/2}
\left( K^{\hat{r}}\,_{i}\,K^{\hat{s}}\,_{j}
- K^{\hat{r}}\,_{j}\,K^{\hat{s}}\,_{i}\right)
\widetilde{E}_{\hat{r}}\,^{j}
\widetilde{E}_{\hat{s}}\,^{i} - h^{1/2}\, R\right]
\nonumber \\
& & \nonumber \\
{\cal H}_{j} & = & \frac{1}{\kappa}\,
D_{k}\left( K^{\hat{r}}\,_{j}
\widetilde{E}_{\hat{r}}\,^{k} - h^{k}_{j}\,
K^{\hat{r}}\,_{i}
\widetilde{E}_{\hat{r}}\,^{i}\right) \label{constraints} \\
& & \nonumber \\
\bar{\cal H}^{\it 1} & = &
\sqrt{\sigma}\left[\gamma\,\varepsilon^{\it 1}
- v \gamma\, (j^{\it 1})_{\vdash}\right] \nonumber\\
& & \nonumber \\
\bar{\cal H}^{\it 1}_{a} & = &
- \sqrt{\sigma}\left[(j^{\it 1})_{a}
- \frac{1}{\kappa}\, \delta_{a} \phi\right]\, .
\nonumber
\end{eqnarray}
Here $R$ is the Ricci scalar of $\Sigma$, and
$\varepsilon^{\it 1}$, $(j^{\it 1})_{\vdash}$,
and $(j^{\it 1})_{b}$
stand for the real parts of the
densities in (\ref{Tset}) and
(\ref{Sset}) (the notation is redundant for
$\varepsilon$ and $j_{\vdash}$,
as these are purely real). Also, at this
stage, the hybrid extrinsic curvature
$K^{\hat{r}}\,_{j}$ is
merely a short-hand
notation for a complicated function of $N$, $V^{j}$,
${\dot{\widetilde{E}\,}\!}_{\hat{r}}\,^{j}$, and
$\widetilde{E}_{\hat{r}}\,^{j}$. Finally,
note that $\bar{\cal H}^{\it 1}$ and
$\bar{\cal H}^{\it 1}_{b}$ are not constraints.

The next step is to calculate the (3+1) form
of the subtraction term,
which can be written as
\begin{equation}
- S^{\it 0} = - \frac{\rm i}{2\kappa}\int_{\cal M}
{\rm d}^{4}x\,\partial_{\rm t}
\left(\epsilon^{\hat{r}\hat{s}\hat{p}}\,
\omega_{\hat{s}\hat{p} j}\,
\widetilde{E}_{\hat{r}}\,^{j}\right)
+ \frac{\rm i}{2\kappa}\int_{\bar{\cal T}}
{\rm d}^{3}x\,\sqrt{- \bar{\gamma}}\,
\epsilon^{\hat{r}\hat{s}\hat{p}}\,
\bar{\tau}_{\hat{s}\hat{p}\hat{r}}\, . \end{equation}
Tedious but straightforward manipulations yield
\begin{eqnarray}
- S^{\it 0} & = &
- \frac{\rm i}{2\kappa}\int_{\cal M}{\rm d}^{4}x\,
\epsilon^{\hat{r}\hat{s}\hat{p}}\,
\omega_{\hat{s}\hat{p} j}\,
{\dot{\widetilde{E}\,}\!}_{\hat{r}}\,^{j} \nonumber \\
& & \hspace{1cm} +
\frac{\rm i}{2\kappa}\int_{\cal M}{\rm d}^{4}x\,\sqrt{h}\,
D_{i}\left(\epsilon^{i}\,_{jk}\,
{\,\,\,}_{_{_{\textstyle \widetilde{}}}}\hspace{-2mm}
E^{\hat{r} j}\,{\dot{\widetilde{E}\,}\!}_{\hat{r}}\,^{k}\right)
+ \frac{\rm i}{2\kappa}\int_{\bar{\cal T}}{\rm d}^{3}x\,
\sqrt{- \bar{\gamma}}\,
\epsilon^{\hat{r}\hat{s}\hat{p}}\,
\bar{\tau}_{\hat{s}\hat{p}\hat{r}}\, ,
\end{eqnarray}
where ${\,\,\,}_{_{_{\textstyle \widetilde{}}}}\hspace{-2mm}
E^{\hat{r} j} = h^{-1/2}\, E^{\hat{r} j}$.
For the middle integral on the right-hand side, we now use
Stokes' theorem for each $\Sigma$ slice and enforce the
radial-gauge condition at the boundary
$B$ of each $\Sigma$. Also, we expand the integrand in the
final integral subject to the
assumption that the $\bar{\cal T}$ triad
is time-gauge with respect to $\bar{u}$. The result of these
calculations is
\begin{eqnarray}
- S^{\it 0} & = &
- \frac{\rm i}{2\kappa}\int_{\cal M}{\rm d}^{4}x\,
\epsilon^{\hat{r}\hat{s}\hat{p}}\,
\omega_{\hat{s}\hat{p} j}\,
{\dot{\widetilde{E}\,}\!}_{\hat{r}}\,^{j}
\nonumber \\
& & \hspace{1cm}
- \frac{\rm i}{2\kappa}\int_{\bar{\cal T}}
{\rm d}^{3}x\,\sqrt{\sigma}\,
\sigma_{bd}\,\epsilon^{\hat{a}\hat{c}}\,
\theta_{[\hat{a}}\,^{b}\,\dot{\theta}_{\hat{c}]}\,^{d}
+  \frac{\rm i}{2\kappa}\int_{\bar{\cal T}}{\rm d}^{3}x\,
\sqrt{- \bar{\gamma}}\, \epsilon^{\bot'\hat{a}\hat{c}}\,
\bar{\tau}_{\hat{a}\hat{c}\bot'}\, . \label{middle}
\end{eqnarray}
Next, ``barring" the last formula in (\ref{2papaya}),
one finds
\begin{equation}
\sigma_{bd}\,
\theta_{[\hat{1}}\,^{b}\,\dot{\theta}_{\hat{2}]}\,^{d}
= \bar{N}\, \bar{\tau}_{\hat{1}\hat{2}\bot'}
+ 1/2\,\epsilon^{ac}\,\delta_{a} \bar{V}_{c} +
\bar{V}^{b}\,\bar{\tau}_{\hat{1}\hat{2} b}\, .
\end{equation}
Insertion of this formula into (\ref{middle}) gives the
desired (3+1) form of the subtraction term,
\begin{equation}
- S^{\it 0} = - \frac{\rm i}{2\kappa}\int_{\cal M}
{\rm d}^{4}x\,
\epsilon^{\hat{r}\hat{s}\hat{p}}\,
\omega_{\hat{s}\hat{p} j}\,
{\dot{\widetilde{E}\,}\!}_{\hat{r}}\,^{j}
- \frac{\rm i}{\kappa}\int_{\bar{\cal T}}
{\rm d}^{3}x\,\sqrt{\sigma}\,
\omega_{\hat{1}\hat{2} b}\, \bar{V}^{b}\, ,
\label{subterm}
\end{equation}
where we have also used
$\bar{\tau}_{\hat{a}\hat{c} b} = \omega_{\hat{a}\hat{c} b}$.

We now turn to the canonical form of the action principle.
We shall avoid the issue of the reality conditions by working
first with canonical form of the real action $S^{\it 1}$.
Therefore, upon adding the pure imaginary boundary term
(\ref{subterm}) to $S^{\it 1}$, we merely introduce a complex
chart on the real phase space.\cite{Ashtekar1} 
The canonical form of the
action $S^{\it 1}$ is the following:
\begin{equation}
S^{\it 1} = \int_{\cal M}{\rm d}^{4}x
\left[ P^{\hat{r}}\,_{j}\,
\dot{\widetilde{E}\,}_{\hat{r}}\,^{j}
- N\,{\cal H} - V^{j}\, {\cal H}_{j}
- 1/2\,\phi_{\hat{r}\hat{s}}\, J^{\hat{r}\hat{s}} \right]
+ \int_{\bar{\cal T}}{\rm d}^{3}x
\left[ - \frac{\phi}{\kappa}\,
\dot{\sqrt{\sigma}\,\,\,\,\,}\!\!\!\!\! -
\bar{N}\, \bar{\cal H}^{\it 1} - \bar{V}^{b}\,
\bar{\cal H}^{\it 1}_{b}\right] .
\label{canonicalpylos}
\end{equation}
In general
$P^{\hat{r}}\,_{j} \neq 1/\kappa\, K^{\hat{r}}\,_{j}$.
Indeed, setting $P_{ij} =
P^{\hat{r}}\,_{j}\, E_{\hat{r} i}$, one has that
\begin{equation}
P_{ij}\,\approx\,\frac{1}{\kappa}K_{ij}\, ,
\end{equation}
where $\approx$ stands for
modulo the {\em rotation constraint},
\begin{equation}
J^{\hat{r}\hat{s}} \equiv
2\, P^{[\hat{r}}\,_{j}\,\delta^{\hat{s}]\hat{t}}\,
\widetilde{E}_{\hat{t}}\,^{j}\, .
\end{equation}
(Note that the Lagrange parameter $\phi_{\hat{r}\hat{s}}$
associated with $J^{\hat{r}\hat{s}}$ is not related to the 
boost parameter $\phi$ and that $\delta^{\hat{r}\hat{s}}$ 
is the Kronecker symbol.) 
Furthermore, in (\ref{canonicalpylos})
$\cal H$, ${\cal H}_{j}$, $
\bar{\cal H}^{1}$, and $\bar{\cal H}^{1}_{b}$ have
the same forms
as given in (\ref{constraints}) but now are built with
$P^{\hat{r}}\,_{j}$ rather than
$1/\kappa\, K^{\hat{r}}\,_{j}$.
In particular, in the canonical action $S^{\it 1}$
\begin{eqnarray}
\bar{\cal H}^{\it 1} & = & \sqrt{\sigma}
\left[\frac{1}{\kappa}\, \gamma\, k
+ v \gamma\, \sigma_{ij}\, P^{ij}\right] \nonumber\\
& &  \\
\bar{\cal H}^{\it 1}_{a} & = & -
\sqrt{\sigma}\left[ n_{i}\,\sigma_{aj}\, P^{ij}
- \frac{1}{\kappa}\, \delta_{a} \phi\right] \nonumber
\end{eqnarray}
(with the radial-gauge condition at the boundary $B$ of
$\Sigma$, one can write
$k = - \omega^{\hat{a}}\,_{\vdash\hat{a}}$).
As is well-known,
the anticommuting Lagrange multiplier $\phi_{\hat{r}\hat{s}}$
associated with the rotation constraint can be geometrically
interpreted as the time component of the connection
forms, \cite{Teitelboim}
\begin{equation}
\phi_{\hat{r}\hat{s}} =
- \Gamma_{\hat{r}\hat{s} {\rm t}} = - \left\langle
\Gamma_{\hat{r}\hat{s}}\, ,\,
\partial/\partial t \right\rangle\, .
\label{rotmultiplier}
\end{equation}
Enforcement of the radial-gauge condition at the boundary
$B$ of each $\Sigma$ slice places a boundary condition on
$\phi_{\hat{r}\hat{s}}$. This boundary
condition is the canonical
version of setting the connection coefficient
$\Gamma_{\hat{a}\vdash \mu}\,\bar{u}^{\mu} = 0$,
where we are working in the RT-gauge described in the
appendix. The coefficient
$\Gamma_{\hat{a}\vdash \mu}\,\bar{u}^{\mu}$
describes the rotation of $n$ as it is parallel
transported along the integral curves of $\bar{u}$.
To see what the required boundary condition on
$\phi_{\hat{r}\hat{s}}$ is, first recall that in
the triad formalism
the vector constraint is not the generator of
diffeomorphisms,
rather
${\cal H}^{\it diff}_{j} = {\cal H}_{j}
- 1/2\, J^{\hat{r}\hat{s}}\,\omega_{\hat{r}\hat{s} j}$
so the vector
constraint generates rotation of the triad.\cite{Teitelboim}
Therefore, the boundary condition
\begin{equation}
- \left. 1/\bar{N}\left(\phi_{\hat{a}\vdash}
+ \omega_{\hat{a}\vdash b}\,
\bar{V}^{b}\right)\right|_{\bar{\cal T}} = 0
\end{equation}
ensures consistency between the selection of the
radial-gauge condition
for the $\Sigma$ triad at $B$ and the evolution of the
triad as obtained
from the variation of the canonical action.

We now add the boundary term (\ref{subterm}) to the canonical
action ({\ref{canonicalpylos}) and get
\begin{equation}
S = \int_{\cal M}{\rm d}^{4}x \left[ {\rm i}\, A^{\hat{r}}\,_{j}\,
\dot{\widetilde{E}\,}_{\hat{r}}\,^{j} -
{\,\,\,}_{_{_{\textstyle \widetilde{}}}}\hspace{-2mm} N\,
\,\,{\,}^{{\,\!}_{{\,\!}^{\textstyle
                        \widetilde{}}}}\hspace{-2.8mm}{\cal H}
- V^{j}\, {\cal H}_{j}
- 1/2\,\phi_{\hat{r}\hat{s}}\, J^{\hat{r}\hat{s}} \right]
+ \int_{\bar{\cal T}}{\rm d}^{3}x\left[
- \frac{\phi}{\kappa}\,
\dot{\sqrt{\sigma}\,\,\,\,\,}\!\!\!\!\! -
\bar{N}\, \bar{\cal H}
- \bar{V}^{b}\, \bar{\cal H}_{b}\right]\, ,
\label{canonicalbassae}
\end{equation}
where in anticipation of dealing with the Ashtekar versions
of the canonical constraints, we have written
${\,\,\,}_{_{_{\textstyle \widetilde{}}}}\hspace{-2mm}
N \equiv h^{-1/2}\, N$ and
$\,\,{\,}^{{\,\!}_{{\,\!}^{\textstyle
                        \widetilde{}}}}\hspace{-2.8mm}{\cal H}
\equiv h^{1/2}\, {\cal H}$. Here
$\bar{\cal H} = \bar{\cal H}^{\it 1}$,
while
$\bar{\cal H}_{b} = \bar{\cal H}^{\it 1}_{b}
+ {\rm i}/\kappa\,\sqrt{\sigma}\,
\omega_{\hat{1}\hat{2} b}$.
{\em Furthermore, for the rest of this
section $A^{\hat{r}}\,_{j}$
is the canonical Ashtekar connection}
\begin{equation}
A^{\hat{r}}\,_{j} = \frac{1}{\kappa}\,\omega^{\hat{r}}\,_{j}
- {\rm i}\, P^{\hat{r}}\,_{j}\, .
\end{equation}
As usual, one may replace the rotation constraint with the
{\em Gauss constraint},
\begin{equation}
J_{\hat{r}} = - \frac{\rm i}{\kappa}\,
^{A} {\cal D}_{j}\,\widetilde{E}_{\hat{r}}\,^{j}
\equiv - \frac{\rm i}{\kappa}\left(D_{j}\,
\widetilde{E}_{\hat{r}}\,^{j} - \kappa
\widetilde{E}_{\hat{s}}\,^{j}\,
\epsilon^{\hat{s}}\,_{\hat{p}\hat{r}}\,
A^{\hat{p}}\,_{j}\right)
= - \frac{1}{2}\,\epsilon_{\hat{r}\hat{s}\hat{t}}\,
J^{\hat{s}\hat{t}}\, ,
\end{equation}
where $^{A} {\cal D}_{j}$ is the derivative operator
associated with the Ashtekar connection.
Moreover, using the Ashtekar curvature,
\begin{eqnarray}
F^{\hat{r}}\,_{jk} & \equiv & 2 D_{[j}A^{\hat{r}}\,_{k]} +
\kappa\,\epsilon^{\hat{r}\hat{s}\hat{t}}\,A_{\hat{s}j}\,
A_{\hat{t}k} \label{2Ashcurvature}\\
  & = &  \frac{1}{\kappa}
\left(2 D_{[j}\omega^{\hat{r}}\,_{k]} +
\epsilon^{\hat{r}\hat{s}\hat{t}}\,\omega_{\hat{s}j}\,
\omega_{\hat{t}k}\right)
- \kappa\,\epsilon^{\hat{r}\hat{s}\hat{t}}\,
P_{\hat{s}j}\,P_{\hat{t}k} - 2 {\rm i}\, E^{\hat{r}}\,_{i}\,
D_{[j}\,P^{i}\,_{k]}\, , \nonumber
\end{eqnarray}
one can build the standard $\Sigma$ constraints:
\begin{eqnarray}
{\cal C} & \equiv & \frac{1}{2}\,
\epsilon^{\hat{r}\hat{s}\hat{t}}\,
\widetilde{E}_{\hat{r}}\,^{i}\,
\widetilde{E}_{\hat{s}}\,^{j}\,F_{\hat{t}ij}
= \,\,{\,}^{{\,\!}_{{\,\!}^{\textstyle
                        \widetilde{}}}}\hspace{-2.8mm}{\cal H}
- {\rm i}\, D_{j}
\left(\widetilde{E}_{\hat{r}}\,^{j}\, J^{\hat{r}}\right)
\nonumber \\
& &  \\
{\cal C}_{j} & \equiv &
i\,\widetilde{E}_{\hat{r}}\,^{i}\,F^{\hat{r}}\,_{ij}
= {\cal H}_{j}
- {\rm i}\,\kappa\, J_{\hat{r}}\, P^{\hat{r}}\,_{j}\, .
\nonumber
\end{eqnarray}
With this machinery, one may rearrange terms in the expression
(\ref{canonicalbassae}) to find
\begin{equation}
S = \int_{\cal M}{\rm d}^{4}x \left\{{\rm i}\,
A^{\hat{r}}\,_{j}\,
{\dot{\widetilde{E}\,}\!}_{\hat{r}}\,^{j}
- {\,\,\,}_{_{_{\textstyle \widetilde{}}}}\hspace{-2mm}
N\, {\cal C} - V^{j}\, {\cal C}_{j}
- \varphi^{\hat{r}}\, J_{\hat{r}} \right\}
+ \int_{\bar{\cal T}}{\rm d}^{3}x
\left[- \frac{\phi}{\kappa}\,
\dot{\sqrt{\sigma}\,\,\,\,\,}\!\!\!\!\! -
{\,\,\,}_{_{_{\textstyle \widetilde{}}}}\hspace{-2mm}
\bar{N}\, \bar{\cal C}
- \bar{V}^{b}\, \bar{\cal C}_{b}\right]\, .
\label{canonical}\end{equation}
The Lagrange multiplier associated with the Gauss constraint
here has the explicit form
\begin{equation}
\varphi^{\hat{r}} = - 1/2\,
\epsilon^{\hat{r}\hat{s}\hat{p}}\,\phi_{\hat{s}\hat{p}}
- {\rm i}\, \delta^{\hat{r}\hat{s}}
\widetilde{E}_{\hat{s}}\,^{j}\,
D_{j} {\,\,\,}_{_{_{\textstyle \widetilde{}}}}\hspace{-2mm}
N + {\rm i} \kappa\,
P^{\hat{r}}\,_{j}\, V^{j}\, .
\end{equation}
Furthermore, now we have
\begin{eqnarray}
\bar{\cal C} & = & \sqrt{h}\, \bar{\cal H}
- {\rm i}\,\gamma\, ({\rm d}r)_{i}\,
\epsilon_{\hat{r}}\,^{\hat{s}\hat{p}}\,
P^{\hat{r}}\,_{j}\,\widetilde{E}_{\hat{s}}\,^{j}\,
\widetilde{E}_{\hat{p}}\,^{i} =
\alpha \sigma
\left(\gamma\,\varepsilon - v \gamma\, j_{\vdash}\right)
\nonumber\\
& & \nonumber \\
\bar{\cal C}_{a} & = & \bar{\cal H}_{a} =
- \sqrt{\sigma}\left[ j_{a}
- \frac{1}{\kappa}\, \delta_{a} \phi\right]\, ,
\nonumber
\end{eqnarray}
where $\alpha = h^{1/2}\,\sigma^{-1/2}$. (Again, $\bar{\cal C}$
and $\bar{\cal C}_{b}$ are not constraints, i.\ e.\ they
do not vanish ``on-shell.")
At this point the densities
$\varepsilon$, $j_{\vdash}$, and $j_{a}$ have the
same forms as in
(\ref{3historyset}) and (\ref{capset}) {\em but are constructed
with the canonical Ashtekar connection}. Therefore, off the
constraint surface in phase space defined by the Gauss
constraint, the energy density $\varepsilon$ is no
longer manifestly
real. Notice that $\bar{\cal C}$ has been defined 
with a higher density 
weight, because it is paired with the boundary
``Lagrange multiplier"
${\,\,\,}_{_{_{\textstyle \widetilde{}}}}\hspace{-2mm}
\bar{N}$, which we have taken as
a density of weight minus one.  We also remark that the
kinematical torsion which is present in the Ashtekar
connection modifies the boost relations. Therefore, for
instance, it is not true that
$\bar{\varepsilon} = \gamma\,\varepsilon
- v \gamma\, j_{\vdash}$
in the canonical picture.

Before considering the
variation of the action (\ref{canonical}),
we find it convenient to rewrite
the Lagrange parameter $\varphi^{\hat{r}}$ in the
following way.
Take
\begin{equation}
\phi_{\hat{r}\hat{s}} =
- \Gamma_{\hat{r}\hat{s}\, {\rm t}} =
- N \Gamma_{\hat{r}\hat{s}\bot}
- \omega_{\hat{r}\hat{s} j}\, V^{j}\, ,
\end{equation}
and also write
\begin{equation}
- {\rm i}\,
\delta^{\hat{r}\hat{s}}
\widetilde{E}_{\hat{s}}\,^{j}\, D_{j}
{\,\,\,}_{_{_{\textstyle \widetilde{}}}}\hspace{-2mm} N
= - {\rm i}\, N\, a^{\hat{r}} =
- {\rm i}\, N \Gamma^{\hat{r}}\,_{\bot\bot} \, ,
\end{equation}
where $a_{\hat{r}} = E_{\hat{r}}[\log N]$ are the
triad components of the
spacetime acceleration of $u$.
With these relations one can set
\begin{equation}
\varphi^{\hat{r}} =
\epsilon^{\hat{r}\hat{s}\hat{p}}\,N
\Gamma^{(+)}\,_{\hat{s}\hat{p}\bot}
- \kappa\, A^{\hat{r}}\,_{j}\, V^{j}\, ,
\end{equation}
which is, of course, essentially the well-known result that
$\varphi^{\hat{r}} =
\epsilon^{\hat{r}\hat{s}\hat{p}}\,
\Gamma^{(+)}\,_{\hat{s}\hat{p} {\rm t}}$.
We shall need the expression for $\varphi^{\vdash}$
when the radial gauge
condition is enforced,
\begin{equation}
\varphi^{\vdash} = 2N\Gamma^{(+)}\,_{\hat{1}\hat{2}\bot}
- \kappa\, A^{\vdash}\,_{j}\, V^{j}\, .
\end{equation}
Using
$V^{j} = V^{\vdash}\, n^{j} + \bar{V}^{b}\,\sigma_{b}^{j}$,
one can put this result in the handy form
\begin{equation}
\varphi^{\vdash} = 2N\Gamma^{(+)}\,_{\hat{1}\hat{2}\bot} -
\kappa\, A^{\hat{r}}\,_{j}\, E_{\hat{r}}\,^{j}\, V^{\vdash} +
{\rm i}\kappa\left(j_{\vdash}\, V^{\vdash} +
j_{c}\,\bar{V}^{c}\right)\, . \label{parameter}
\end{equation}

Direct calculation yields the following for the
variation of the
canonical action (\ref{canonical}):
\begin{eqnarray}
\delta S & = & \left({\rm terms\,\,which\,\,give\,\,the\,\,
constraints\,\,and\,\,
equations\,\,of\,\,motion}\right) + {\rm i}\int_{t'}^{t''}
{\rm d}^{3}x\, A^{\hat{r}}\,_{j}\,\delta
\widetilde{E}_{\hat{r}}\,^{j}
\nonumber \\
&  & - \int_{\bar{\cal T}}{\rm d}^{3}x\,\left[
{\,\,\,}_{_{_{\textstyle \widetilde{}}}}\hspace{-2mm}
\bar{\cal C}\,\delta\bar{N} +
\bar{\cal C}_{b}\,\delta \bar{V}^{b}
- \sqrt{\sigma}\bar{N}/2\left(\gamma\, s^{ab} - v\gamma\,
t^{ab} + 1/\kappa\,
\bar{u}[\phi]\,\sigma^{ab} +
\Delta\,\sigma^{ab}\right)(\delta\theta)_{ab}\right]
\nonumber \\
& &  + \frac{1}{\kappa}\int_{\bar{\cal T}}
{\rm d}^{3}x\,\sqrt{\sigma}
\left[\kappa \bar{N}
\left(\gamma\, j_{\vdash} - v\gamma\,\varepsilon\right)
+ \delta_{a}\,\bar{V}^{a}
- 1/2\,\sigma^{ab}\,\dot{\sigma}_{ab}\right]\delta \phi
- \frac{1}{\kappa}\int^{B''}_{B'}
{\rm d}^{2}x\,\phi\,\delta \sqrt{\sigma}\, ,
\end{eqnarray}
where here $\bar{N}$ is $h^{1/2}
{\,\,\,}_{_{_{\textstyle \widetilde{}}}}\hspace{-2mm}
\bar{N}$,
and it would
perhaps be better to express $(\delta\theta)_{ab}$
as a variation
in terms of the
densitized dyad (as is certainly possible). Also above,
\begin{equation}
\bar{N}\Delta =  {\rm i}\,
A^{\hat{r}}\,_{j}\, E_{\hat{r}}\,^{j}\,
V^{\vdash} + j_{\vdash}\, V^{\vdash} + j_{c}\,\bar{V}^{c}
+ {\rm i}/\kappa
\left(\varphi^{\vdash} - 
2 N\,\Gamma^{(+)}\,_{\hat{1}\hat{2}\bot}\right)\, .
\end{equation}
Let us verify that our result for the variation of the canonical
action agrees with the variation (\ref{delphi}) of the 
non-canonical action. With the interpretation (\ref{parameter}) 
$\Delta$ vanishes. Therefore, enforcing the Gauss constraint and 
using the boost relations (\ref{boostlaws}), one finds that
\begin{eqnarray}
\left(\delta S\right)_{\bar{\cal T}} & \approx  &
- \int_{\bar{\cal T}}{\rm d}^{3}x\,\sqrt{\sigma}
\left[\bar{\varepsilon}\,\delta\bar{N}
- \bar{\jmath}_{b}\,\delta \bar{V}^{b}
- \bar{N}/2\,\,\bar{s}^{ab}\,(\delta\theta)_{ab}\right]
\nonumber \\
& &  + \frac{1}{\kappa}\int_{\bar{\cal T}}{\rm d}^{3}x\,
\sqrt{\sigma}\left(\kappa \bar{N}
\,\bar{\jmath}_{\vdash'} + \delta_{a}\,\bar{V}^{a}
- 1/2\,\sigma^{ab}\,\dot{\sigma}_{ab}\right)\delta \phi
- \frac{1}{\kappa}\int^{B''}_{B'}
{\rm d}^{2}x\,\phi\,\delta \sqrt{\sigma}\, ,
\end{eqnarray}
where now one must again consider the quasilocal densities
to be expressed in terms of the Sen connection. The density
$\bar{\jmath}_{\vdash'} =
- 1/\kappa\, \sigma^{ab}\,\bar{l}_{ab}$,
and in the non-canonical picture
\begin{equation}
2 \bar{N} \bar{l}_{ab} = - \dot{\sigma}_{ab} +
\delta_{a} \bar{V}_{b} + \delta_{b} \bar{V}_{a}\, ,
\end{equation}
so the
middle integral on the right-hand side vanishes in this
case. This means that $\phi$ is {\em not} held fixed in the
canonical variational principle, as the equations of motion
ensure that this term vanishes for arbitrary variations 
$\delta \phi$ about a classical solution. Using the barred
versions of (\ref{2historyset}), we then have
\begin{equation}
\left(\delta S\right)_{\bar{\cal T}} \approx
{\rm i}\int_{\bar{\cal T}}{\rm d}^{3}x\,
\bar{\cal A}^{\hat{r}}\,_{j}\,\delta
\left(\sqrt{-\bar{\gamma}}\,
\bar{\xi}_{\hat{r}}\,^{j}\right)
- \frac{1}{\kappa}\int^{B''}_{B'}
{\rm d}^{2}x\,\phi\,\delta \sqrt{\sigma}\, ,
\end{equation}
in agreement with the variation (\ref{delphi}) of the
non-canonical action.

\section{Discussion}
We conclude with (i) a description
of some of the new developments
in the theory of quasilocal stress-energy-momentum
which will appear in Ref.\ \cite{BYL}.
We also briefly comment on several technical matters.
These are (ii) the issue of additional subtraction-term 
contributions to the action,
(iii) the relationship of our formalism with the Sparling
two-forms, and (iv) a problem encountered in the attempt to 
extend the Brown-York notion of gravitational charge to the
Ashtekar-variable version of the theory.

(i) Since much of the analysis in this paper is based on
Ref.\ \cite{BYL}, it is appropriate to describe a few results
which will be found in this upcoming work. First,
Ref.\ \cite{BYL} deals exclusively with the metric-variable version
of quasilocal stress-energy-momentum, though this is not a
distinction between Ref.\ \cite{BYL} and the
present paper that we
wish to highlight in this paragraph. 
Regardless of the choice of gravitational
variables, the results to be found in
Ref.\ \cite{BYL} are {\em more}
general than those presented here in the
following sense. In this
paper the set
$\{\varepsilon, j_{\vdash}, j_{a}, s^{ab} , t^{ab}\}$
of quasilocal densities is associated with a family of Eulerian
(or surface-forming) observers at $B$. We have derived the
transformations rules between two different sets of quasilocal
densities, but each of the sets is associated with its own family
of Eulerian observers. Ref.\ \cite{BYL} also considers a set of
densities
$\{\varepsilon, j_{\vdash}, j_{a}, s^{ab} , t^{ab}\}$
(but built with metric variables). However, in Ref.\ \cite{BYL}
the densities need not be associated
with a family of Eulerian
observers. That is, they may describe the stress-energy-momentum
content of the gravitational field which is associated with a
family
of Lagrangian (or non-surface forming) observers, such as those
determined by the timelike Killing field of the Kerr geometry.
The transformation rules between the Lagrangian
set of densities and a set associated with an arbitrary family of
Eulerian observers will be given. Hence, the boost relations
which will appear in Ref.\ \cite{BYL} are more general than those
appearing here (the Eulerian-Eulerian boost relations arise as a
special case). It may possible to derive these more general boost
relations in the Ashtekar-variable framework as well, however,
such a derivation would be considerably more difficult from a 
technical standpoint.

(ii) As mentioned in the introduction, we have chosen to formally
treat the imaginary boundary term $- S^{\it 0}$ as a true
subtraction term \`{a} la Brown and York. However, we now argue
that in some contexts it is necessary to consider the freedom to
append to the action $S = S^{\it 1} - S^{\it 0}$ an additional
subtraction term $- S'^{\it 0}$.
In the interest of economy we restrict our argument to matters
concerning the quasilocal energy surface density $\varepsilon$,
though perhaps much of this discussion also pertains to the 
other quasilocal
densities. Consider first the Brown-York expression
\begin{equation}
\varepsilon = \frac{1}{\kappa}\left( k - k^{\it 0}\right)\, .
\end{equation}
In the metric formalism, as in this paper, $k$ represents the
trace of the extrinsic curvature of $B$ as
embedded in $\Sigma$ and comes from a base
action in the derivation. The
$k^{\it 0}$ term represents the trace of the extrinsic curvature
of a two-surface which has the same metric as $B$, but which is
uniquely embedded in a three-dimensional manifold possibly
different than
$\Sigma$. In the Brown-York formalism it arises from a {\em real}
subtraction-term contribution to the
action. When possible, this real subtraction term is typically chosen 
such that the different three-space is $R^{3}$, and hence the 
$k^{\it 0}$ term references the energy against flat-space. 
For a given asymptotically-flat spacetime, the presence of 
the appropriate $k^{\it 0}$ term is crucial if the quasilocal 
energy,
\begin{equation}
E = \frac{1}{\kappa}\int_{B}{\rm d}^{2}x\,
\sqrt{\sigma}\left(k - k^{\it 0}\right)\, ,
\end{equation}
is to agree with the ADM notion of energy in the
suitable limit.\cite{BMYW,BY,Martinez}

Though we have added the imaginary boundary term $- S^{\it 0}$
to $S^{\it 1}$ in this paper, the resulting quasilocal energy,
\begin{equation}
E = \int_{B}{\rm d}^{2}x\,\sqrt{\sigma}\,\epsilon^{\hat{a}\hat{c}}\,
A_{\hat{a} b}\,\theta_{\hat{c}}\,^{b} \approx
\frac{1}{\kappa}\int_{B}{\rm d}^{2}x\,
\sqrt{\sigma}\, k\, ,
\end{equation}
is really only the ``unreferenced" energy. (As we have seen,
the particular subtraction term used in this work makes no
contribution to $\varepsilon$ and thus $E$.) If we wish to put
the Ashtekar-variable expression for the quasilocal energy
into full accord with the ADM notion of energy, then we
should allow for the freedom to append to the action {\em yet
another} subtraction term $- S'^{\it 0}$. Use of the more general
action $S' = S'^{\it 1} - S'^{\it 0} = S^{\it 1} - S^{\it 0}
- S'^{\it 0}$ in our analysis would yield
\begin{equation}
E = \int_{B}{\rm d}^{2}x\,\sqrt{\sigma}\,
\epsilon^{\hat{a}\hat{c}}\,
\theta_{\hat{c}}\,^{b}
\left( A_{\hat{a} b} - A^{\it 0}_{\hat{a} b} \right)
\end{equation}
for the Ashtekar-variable quasilocal energy. The new
reference-point contribution
$\epsilon^{\hat{a}\hat{c}}\, A^{\it 0}_{\hat{a} b}\,
\theta_{\hat{c}}\,^{b}$
stems from
$- S'^{\it 0}$. At this point $- S'^{\it 0}$
is an arbitrary functional of $\bar{\cal T}$ data.
With this new freedom, the quasilocal energy can be defined to
agree with the ADM energy for asymptotically-flat spacetimes 
(in the suitable limit that $B$ becomes the two-sphere at infinity).
A fuller discussion of this issue will appear elsewhere.

(iii) The real and complex Sparling
two-forms obey the {\em Sparling relation}
\begin{equation}
{\rm d} \sigma_{\hat{\mu}} = {\rm d}
\sigma^{(+)}\,_{\hat{\mu}} =
\tau_{\hat{\mu}} + G_{\hat{\mu}}\,^{\hat{\sigma}}\,
e^{*}_{\hat{\sigma}}\, ,
\end{equation}
where $e^{*}_{\hat{\sigma}}$ is a basis for three-forms,
$G_{\hat{\mu}}\,^{\hat{\sigma}}$ is the Einstein tensor, and
$\tau_{\hat{\mu}}$ are the {\em Sparling
three-forms}. The explicit form for $\tau_{\hat{\mu}}$
(which is real) is not needed here but may be found in,
for example, Ref.\ \cite{Sparling}. The Sparling relation
suggests that $\tau_{\hat{\mu}}$ (when pulled-backed to a
three-dimensional slice $\Sigma$ of spacetime)
may be interpreted as a tetrad-dependent local energy-momentum
density for the gravitational field.\cite{Sparling,Goldberg}
The corresponding frame-dependent potential can be taken
either as $\sigma_{\hat{\mu}}$ or $\sigma^{(+)}\,_{\hat{\mu}}$.
If we fix a two-surface $B$ and its spanning
three-slice $\Sigma$ in spacetime, then the boundary structure
of our selection provides a natural (almost-unique) frame at
$B$. Namely, the radial time-gauge tetrad
of {\em Appendix A}, which has the $\Sigma$ hypersurface normal
$u$ as its time leg and $n$, the normal of $B$ in $\Sigma$,
as its third space leg. With this frame choice,
the pullbacks $s^{*}(\sigma^{(+)}\,_{\hat{\mu}})$
to $B$ ($s$ is the inclusion mapping
$s: B \rightarrow {\cal M}$) are the following:
\begin{eqnarray}
s^{*}(\sigma^{(+)}\,_{\bot}) & = &
- \kappa\, \varepsilon\, \sqrt{\sigma}\, {\rm d}^{2}x
\nonumber \\
s^{*}(\sigma^{(+)}\,_{\vdash}) & = & \kappa\, j_{\vdash}\,
\sqrt{\sigma}\, {\rm d}^{2}x \\
s^{*}(\sigma^{(+)}\,_{\hat{a}}) & = &
\kappa\, \theta_{\hat{a}}\,^{b}\,
j_{b}\, \sqrt{\sigma}\, {\rm d}^{2}x \, .\nonumber
\end{eqnarray}
Here these are expressed in terms of the $\Sigma$ Sen
connection and triad. Further,
the pullbacks $s^{*}(\sigma_{\hat{\mu}})$ of the real Sparling
two-forms are just the real parts of the above expressions.
But only the last expression is complex, so
$s^{*}(\sigma_{\bot}) = s^{*}(\sigma^{(+)}\,_{\bot})$
and $s^{*}(\sigma_{\vdash}) = s^{*}(\sigma^{(+)}\,_{\vdash})$.
One has $s^{*}(\sigma_{\hat{a}}) =
\kappa\,\theta_{\hat{a}}\,^{b} (j^{\it 1})_{b}\,
\sqrt{\sigma}\, {\rm d}^{2} x$.
See Ref.\ \cite{Lau} for more details.

(iv) The Brown-York notion of gravitational charge is based
on the $\bar{\cal T}$ momentum constraint,
\begin{equation}
- 2\, \bar{\cal D}_{i}
\left( \bar{\pi}^{i}\,_{j}
- (\bar{\pi}^{\it 0})^{i}\,_{j}\right)
= - 1/\kappa\,
\sqrt{- \bar{\gamma}}\,\bar{\gamma}^{\mu}_{j}\,
\bar{n}^{\lambda}\,
G_{\mu\lambda}\, ,
\end{equation}
where $\bar{\pi}^{ij}$ is given in (\ref{momenta}) and,
in the metric
formalism, $(\bar{\pi}^{\it 0})^{ij}$
depends only on $\bar{\gamma}_{ij}$ (and
so it is annihilated by $\bar{\cal D}_{i}$). Now we work
on-shell and in vacuum,
so this expression vanishes.  Brown and York define
a ``stress tensor"
$\bar{\tau}^{ij} = 2/\sqrt{-\bar{\gamma}}\left( \bar{\pi}^{ij}
- (\bar{\pi}^{\it 0})^{ij}\right)$.
Assume that $\bar{\cal T}$ possesses a Killing
field $\zeta^{j}$,
and so
$\bar{\cal D}_{i}\,\bar{\tau}^{ij}\,\zeta_{j} = 0$.
Therefore, since
$-\bar{u}_{i}\,\bar{\tau}^{ij}
= \bar{\varepsilon}\, \bar{u}^{j}
+ \bar{\jmath}_{b}\,\sigma^{b j}$,
one has the following conserved charge:\cite{BY}
\begin{equation}
Q_{\zeta}(B) = \int_{B} {\rm d}^{2}x\,
\sqrt{\sigma}\left(\bar{\varepsilon}\, \bar{u}^{j}
+ \bar{\jmath}_{b}\,\sigma^{b j}\right)\zeta_{j}\, .
\end{equation}
When attempting to introduce such a notion of charge into
our formalism, we run into some
difficulty since the subtraction term $S'^{\it 0}$
may be triad-dependent. (Here $S'^{\it 0}$ may or may not represent the
particular subtraction term $S^{\it 0}$ considered in this
work.) The natural way around this difficulty
is the following. First define
\begin{equation}
(\bar{\Pi}^{\it 0})^{\hat{r}}\,_{j} =
\delta S'^{\it 0}/\delta
\left(\sqrt{-\bar{\gamma}}\,\bar{\xi}_{\hat{r}}\,^{j}\right)\, .
\end{equation}
In our situation $(\bar{\Pi}^{\it 0})_{ij}
= \bar{\xi}_{\hat{r} i}\,(\bar{\Pi}^{\it 0})^{\hat{r}}\,_{j}$
is not necessarily annihilated by $\bar{\cal D}_{j}$,
though $\bar{\cal D}_{j}\, (\bar{\Pi}^{\it 0})_{(ik)} = 0$.
Therefore, set $(\bar{\pi}^{\it 0})^{ij}
= \sqrt{-\bar{\gamma}}/2
\left((\bar{\Pi}^{\it 0})^{k}\,_{k}\,\bar{\gamma}^{ij}
- (\bar{\Pi}^{\it 0})^{(ij)}\right)$ and use it in the
above construction. The charge $Q_{\zeta}$ may now be
complex, but, subject to the assumptions made above, it is
conserved.

\section{Acknowledgments}
For encouragement and a number of discussions it is a pleasure to
thank J.\ D.\ Brown and J.\ W.\ York.
J.\ D.\ Brown provided me with some unpublished notes
which became the basis for some parts of this work.
I also thank S.\ Sinha for helpful remarks.
This research has been supported in part by the Fonds zur 
F\"{o}r\-der\-ung der wis\-sen\-schaft\-lich\-en For\-schung (Lise Meitner Fellowship
M-00182-PHY), by a post-doctoral fellowship from the Indo-US Exchange Programme, and by the University Grants Commission of India.
Also, much of this work was done at the University of North
Carolina at Chapel Hill
with support provided by the National Science Foundation,
grant number PHY-8908741.

\appendix
\begin{center}
\bf APPENDIX: KINEMATICAL FRAMEWORK
\end{center}
{\em Appendices A}, {\em B}, and {\em C} outline a kinematical
framework for examining how the intrinsic and extrinsic geometry
of spacetime as foliated by a family $\bar{\cal T}$
hypersurfaces is related to the intrinsic and
extrinsic geometry of spacetime as foliated by a family of $\Sigma$
hypersurfaces. With this
framework one can express objects such as
the $\bar{\cal T}$ extrinsic
curvature $\bar{\Theta}_{ij}$ or the time-gauge
$\bar{\cal T}$ Sen connection
$\bar{{\cal A}}^{\hat{r}}\,_{j}$ in terms of the
intrinsic and extrinsic geometry of spacetime as
foliated by $\Sigma$
hypersurfaces. Such a ``splitting" of the
$\bar{\cal T}$ Sen connection is
needed in order to derive a similar splitting of
the
$\bar{\Sigma}$ radial-gauge Sen connection
$\bar{A}^{\hat{r}}\,_{j}$ in
terms of the geometry of the $\Sigma$ foliation.
The splitting of
$\bar{A}^{\hat{r}}\,_{j}$ is
used to obtain the boost laws
(\ref{boostlaws}) for the quasilocal
densities. The kinematical framework consists of (i) two distinct
spacetime tetrads (one adapted to the $\Sigma$ foliation and one
adapted to the $\bar{\cal T}$ foliation), (ii) the
transformation equations
between these tetrads, and (iii) the inhomogeneous transformation
law between the sets of associated connection coefficients. The
relevant spacetime tetrads are constructed in {\em Appendix A},
and their associated connection coefficients are tabulated in
{\em Appendix B}. {\em Appendix C} outlines the
splitting procedure
by applying it to $\bar{\Theta}_{ij}$, the simplest example.
We then
quote the splitting results for the $\bar{\cal T}$
time-gauge connection
coefficients $\bar{\tau}_{\hat{r}\hat{s} j}$,
$\bar{{\cal A}}^{\hat{r}}\,_{j}$,
and $\bar{A}^{\hat{r}}\,_{j}$. The final
{\em Appendix D} applies some
of this formalism to explain the origin of the
corner terms in
the action (\ref{cico}).

\section{Adapted tetrads} The boundary structure of
${\cal M}$ suggests
two natural classes of spacetime tetrads.
The first class is a
subclass of time-gauge
tetrads determined by the boundary structure of $\Sigma$.
The second
class is a subclass of ``radial-gauge" tetrads
determined by the
$B$ foliation of $\bar{\cal T}$.
These tetrads need only be defined on
some small spacetime neighborhood surrounding a
portion of $\bar{\cal T}$. We do not
address the issue of whether or not either of these tetrads can
be extended globally over all of ${\cal M}$.

\subsection{Radial time-gauge tetrads}
Enforcement of the {\em time gauge} condition locks the time leg
of the tetrad to the $\Sigma$ foliation normal
$u$. This condition
is indicated by replacing the tetrad time label $\hat{0}$ with
$\bot$ so that $e_{\bot} = u$. Because each $\Sigma$ slice has a
boundary $B$, a natural subclass of all time-gauge tetrads exists
which is determined by an auxiliary condition on
$\bar{\cal T}$.  This
further requirement is that
{\em on the three-boundary} $\bar{\cal T}$
one of the space legs of the tetrad, chosen to be
$e_{\vdash} \equiv e_{\hat{3}}$, coincides with $n$. One should
note that this correspondence is not made between $e_{\hat{3}}$
and $\bar{n}$ in general. Such a choice of
tetrad is said to obey the
{\em radial time-gauge} or {\em RT-gauge} (see Figure 
(\ref{u-and-nfig})). RT-gauge indices and
labels take the values $(\bot,\hat{1},\hat{2},\vdash)$. Now the
usual assumption is that the vector field $\partial/\partial t$
points everywhere tangent to the hypersheets of constant $r$.
Equivalently,
$\left\langle {\rm d}r, \partial/\partial t\right\rangle = 0$
or $\partial r/\partial t = 0$,
and the $r$ coordinate is Lie transported along the integral
curves of the time vector field. This assumption results in
almost no loss of physical generality. It does demand that the
integral curves of the time vector field may not emerge from
or flow into the three-boundary $\bar{\cal T}$.
However, since the
spacetime-filling extension of the three-boundary
$\bar{\cal T}$ in
terms of hypersheets of constant $r$ is
completely arbitrary,
on the interior of $\Sigma$ these integral curves
can be chosen to
flow in any direction (as long as the tangent field
$\partial/\partial t$ lies at each point within the future
light-cone).  Subject to the requirement
$\partial r/\partial t = 0$, one can write the most general
{\em radial vector field} mapped to unity by
${\rm d}r$ as
$\partial /\partial r = \alpha\, n + \beta$,
which is similar to the familiar formula
$\partial/\partial t = N\, u + V$.
As seen earlier, the variables $\alpha$ and $\beta^{a}$ are
respectively the kinematical ``lapse" and ``shift"
associated with the induced radial foliation of the
$\Sigma$ slices.
Therefore, we have the following explicit formulae
for the
RT-gauge tetrad and cotetrad:
\begin{eqnarray}
e_{\bot} = \frac{1}{N}\left(\frac{\partial}{\partial t}
- V^{\hat{a}}\, E_{\hat{a}} - V^{\vdash}\, E_{\vdash} \right)
& \hspace{1cm} &
e^{\bot} = N{\rm d}t \nonumber \\
& & \nonumber \\
e_{\hat{a}} = E_{\hat{a}} = \theta_{\hat{a}} & & e^{\hat{a}} =
\theta^{\hat{a}} + V^{\hat{a}}{\rm d}t
+ \beta^{\hat{a}}{\rm d}r \hspace{10mm} \label{pineapple}\\
& &\nonumber \\
e_{\vdash} = E_{\vdash} =
\frac{1}{\alpha}\left(\frac{\partial}{\partial r} -
\beta^{\hat{a}}\,\theta_{\hat{a}} \right)
& &
e^{\vdash} = \alpha\, {\rm d}r + V^{\vdash}{\rm d}t\, .
\nonumber
\end{eqnarray}

\subsection{Time radial-gauge tetrads}
The {\em radial-gauge condition} requires that one of the space
legs of the tetrad, taken to
be $e_{\vdash'} \equiv e_{\hat{3}}$ , coincides with the
$\bar{\cal T}$ normal $\bar{n}$.
A natural further requirement can be placed
on radial-gauge tetrads. Namely, the time leg
$e_{\bot'} \equiv e_{\hat{0}}$ can be tied to the $B$ timelike
normal $\bar{u}$, so the indices and labels associated with this
class of tetrads run over
$(\bot' , \hat{1}, \hat{2}, \vdash')$.
Such a tetrad is referred as {\em time radial-gauge} or
{\em TR-gauge} (see Figure (\ref{u-and-nfig})). 
Now the radial vector field is written as
$\partial/\partial r = \bar{\alpha}\, \bar{n} + \bar{\beta}$,
though it still points tangent to the $\Sigma$
slices. The variables
$\bar{\alpha}$ and $\bar{\beta}^{\alpha}$ are associated with the
$\bar{\cal T}$ foliation of ${\cal M}$. On $\bar{\cal T}$
one can express the time vector field as
$\partial/\partial t = \bar{N} \bar{u} + \bar{V}$,
where $\bar{N}$ and $\bar{V}^{a}$ are the gauge variables
associated with the $B$ foliation of $\bar{\cal T}$.
The RT-gauge tetrad and cotetrad is
\begin{eqnarray}
e_{\bot'} = \bar{\xi}_{\bot'} =
\frac{1}{\bar{N}}\left(\frac{\partial}{\partial t} -
\bar{V}^{\hat{a}}\, \theta_{\hat{a}}\right)
& \hspace{1cm} &
e^{\bot'} =
\bar{N} {\rm d}t + \bar{\beta}^{\bot'} {\rm d}r \nonumber \\
& & \nonumber \\
e_{\hat{a}} = \bar{\xi}_{\hat{a}} = \theta_{\hat{a}}
& \hspace{1cm} &
e^{\hat{a}} = \theta^{\hat{a}} + \bar{V}^{\hat{a}}{\rm d}t
+ \bar{\beta}^{\hat{a}}{\rm d}r \hspace{10mm}\label{pea} \\
& & \nonumber \\
e_{\vdash'} =
\frac{1}{\bar{\alpha}}\left(\frac{\partial}{\partial r} -
\bar{\beta}^{\hat{a}}\, \bar{\xi}_{\hat{a}} -
\bar{\beta}^{\bot'}\, \bar{\xi}_{\bot'} \right)
& & e^{\vdash'} = \bar{\alpha} {\rm d}r\, .
\nonumber
\end{eqnarray}
We explain the chosen notation further
in the next two paragraphs.

\section{Associated connection coefficients} For the special
tetrads considered above, certain of the corresponding connection
coefficients have notable geometric meanings. This subsection is
a glossary of various connection coefficients and their
geometric interpretations. RT-gauge
connection coefficients are represented
as $\Gamma^{\hat{\rho}}\,_{\hat{\sigma}\hat{\mu}}$, while
TR-gauge connection
coefficients are represented by
$\Gamma^{\hat{\rho}'}\,_{\hat{\sigma}'\hat{\mu}'}$.
Note that inspection of the indices allows one to
discern which set of connection coefficients is being dealt with.

Perhaps a few more comments on the notational scheme will be
clarifying for the reader. The RT-gauge tetrad
$e_{\hat{\mu}}$ and
TR-gauge tetrad $e_{\hat{\mu}'}$ are both tetrads
on the same spacetime $\cal M$. As has been evident, $``e"$
is used to denote both tetrads,
and it is the type of label (primed or unprimed)
carried by $``e"$
which makes the notational
distinction between the two tetrads. Clearly then, though the
$\Gamma^{\hat{\rho}}\,_{\hat{\sigma}\hat{\mu}}$ and
$\Gamma^{\hat{\rho}'}\,_{\hat{\sigma}'\hat{\mu}'}$
are different sets of connection coefficients, they specify
the {\em same} spacetime connection
(that of Levi Civita); and so
we use $\Gamma$ for both, again
letting the labels make the notational distinction between the
two sets.  Notice that neither $e^{\bot'}$ nor $e^{\vdash}$ need be
surface forming. For instance, $e^{\bot'} \wedge {\rm d}
e^{\bot'} \neq 0$ in general. Technically put, in general 
$e^{\bot'}$ and $e^{\vdash}$ will not satisfy the Fr\"{o}benius
condition. Therefore, $e^{\bot'}$ is
not necessarily of the form $\bar{N} {\rm d}\bar{t}$, where
$\bar{t}$ is a coordinate which specifies some 
$\bar{\Sigma}$ slices. However, this is unimportant for our 
calculations. What is important, is that the clamped 
$\bar{\Sigma}$ slices have a future-pointing normal $\bar{u}$ 
which agrees with $e_{\bot'}$ {\em at the physical boundary} 
$\bar{\cal T}$. In fact, the $\bar{\Sigma}$ slices are
determined by this condition, but clearly their extensions
off the three-boundary are highly non-unique. Ultimately, we are 
interested in a $\Sigma$ 3-slice and $\bar{\Sigma}$ 3-slice which 
span the same $B$ 2-slice of the timelike boundary $\bar{\cal T}$.
We want to compare the Cauchy data of the two slices {\em at the 
mutually bounding two-surface} $B$. Our formalism handles this 
issue but does not assume that the $\bar{\Sigma}$ normal $\bar{u}$ 
agrees with $e_{\bot'}$ off the boundary.

In the lists of this appendix,
since the geometry of ${\cal M}$ is torsion-free
(i.\ e.\ the torsion
two-form of Cartan vanishes
\cite{DiffGeom}), all of the extrinsic curvature tensors are
symmetric. Note that for the extrinsic
curvature tensors defined below, we adopt a different convention
for the staggering of indices than the convention used in Refs.
\cite{BY,BYL}. However, since all of these tensors are symmetric, all of our
results match those found in  Refs. \cite{BY,BYL}.

\subsection{RT-gauge connection coefficients}
The RT-gauge connection coefficients are tailored to $B$
as embedded in $\Sigma$. We have the following correspondences:

\begin{eqnarray}
K^{\hat{s}}\,_{\hat{r}} & \equiv &
- e^{\hat{s}}\,_{\mu}\, e_{\hat{r}}\,^{\nu}\, \nabla_{\nu}\,
e_{\bot}\,^{\mu}  =
-\,\Gamma^{\hat{s}}\,_{\bot\hat{r}} \nonumber \\
& & \nonumber \\
a^{\hat{r}} & \equiv &
e^{\hat{r}}\,_{\mu}\, e_{\bot}\,^{\nu}\, \nabla_{\nu}\,
e_{\bot}\,^{\mu} = \Gamma^{\hat{r}}\,_{\bot\bot} \nonumber \\
& & \label{RTcc} \\
k^{\hat{a}}\,_{\hat{c}} & \equiv & -  e^{\hat{a}}\,_{\mu}\,
e_{\hat{c}}\,^{\nu}\, \nabla_{\nu}\, e_{\vdash}\,^{\mu}
 = -\Gamma^{\hat{a}}\,_{\vdash\hat{c}}  =
- \omega^{\hat{a}}\,_{\vdash\hat{c}}
\nonumber \\
& & \nonumber \\
b^{\hat{r}} & \equiv &  e^{\hat{r}}\,_{\mu}\,
e_{\vdash}\,^{\nu}\, \nabla_{\nu}\, e_{\vdash}\,^{\mu}
= \Gamma^{\hat{r}}\,_{\vdash\vdash}\, , \nonumber
\end{eqnarray}
Note that the formulas for $K^{\hat{r}}\,_{\hat{s}}$
and $a^{\hat{r}}$
are general time-gauge expressions. Also,
$b^{\hat{r}}$ are the tetrad components of the spacetime
``acceleration" of $n$, while the $\Sigma$ ``acceleration"
of $n$ has components $b^{\hat{a}}$. For $b^{\hat{r}}$
the $\hat{r}$ is a ${\cal T}$ index and can take the values
$(\bot , \hat{a})$, while for
$K^{\hat{r}}\,_{\hat{s}}$ and $a^{\hat{r}}$ the
$\hat{r}$ and $\hat{s}$ are
$\Sigma$ indices taking the values
$(\hat{1},\hat{2},\vdash)$.

\subsection{TR-gauge connection coefficients}
The TR-gauge connection coefficients are tailored to $B$
as embedded in $\bar{\cal T}$. We have the following
correspondences:
\begin{eqnarray}
\bar{\Theta}^{\hat{r}}\,_{\hat{s}} & \equiv &
- e^{\hat{r}}\,_{\mu}\, e_{\hat{s}}\,^{\nu}\,
\nabla_{\nu}\, e_{\vdash '}\,^{\mu} =
- \Gamma^{\hat{r}}\,_{\vdash'\hat{s}} \nonumber \\
& & \nonumber \\
\bar{b}^{\hat{r}} & \equiv & e^{\hat{r}}\,_{\mu}\,
e_{\vdash'}\,^{\nu}\, \nabla_{\nu}\,
e_{\vdash'}\,^{\mu} =
\Gamma^{\hat{r}}\,_{\vdash'\vdash'} \nonumber \\
& & \label{TRcc} \\
\bar{l}^{\hat{a}}\,_{\hat{c}} & \equiv & -
e^{\hat{a}}\,_{\mu}\, e_{\hat{c}}\,^{\nu}\,
\nabla_{\nu}\, e_{\bot'}\,^{\mu}
= - \Gamma^{\hat{a}}\,_{\bot'\hat{c}} = -
\bar{\tau}^{\hat{a}}\,_{\bot'\hat{c}} \nonumber \\
& & \nonumber \\
\bar{a}^{\hat{r}} & \equiv &  e^{\hat{r}}\,_{\mu}\,
e_{\bot'}\,^{\nu}\, \nabla_{\nu}\, e_{\bot'}\,^{\mu}
= \Gamma^{\hat{r}}\,_{\bot'\bot'}\, . \nonumber
\end{eqnarray}
Like before, the formulas for
$\bar{\Theta}^{\hat{r}}\,_{\hat{s}}$
and $\bar{b}^{\hat{r}}$ are general radial-gauge expressions.
For $\bar{\Theta}^{\hat{r}}\,_{\hat{s}}$ and
$\bar{b}^{\hat{r}}$
in this list the $\hat{r}$ and $\hat{s}$
are $\bar{\cal T}$ indices taking the values
$(\bot' , \hat{a})$.
The $\bar{a}^{\hat{r}}$ ($\hat{r}$ can take the values
$(\hat{a}, \vdash')$) are the tetrad components of
the spacetime acceleration of $\bar{u}$, while the
$\bar{\cal T}$ acceleration of $\bar{u}$ has
components
$\bar{a}^{\hat{c}} = \bar{\tau}^{\hat{c}}\,_{\bot'\bot'}$.

\section{Splitting procedure}

\subsection{Transformation equations}
The set (\ref{set}) of
transformations for the metric variables can be used to express
the transformations between the RT-gauge tetrad
(\ref{pineapple}) and the TR-gauge tetrad (\ref{pea}). For
example,
\begin{eqnarray}
e_{\bot'} & = &
\frac{1}{\bar{N}}\left( \frac{\partial}{\partial
t} - \bar{V}^{b}\, \frac{\partial}{\partial x^{b}}\right)
\nonumber \\ & = &
\frac{\gamma}{N}\left(\frac{\partial}{\partial t}
- V^{b}\, \frac{\partial}{\partial x^{b}} -
V^{\rm r}\, \beta^{b}\,\frac{\partial}{\partial x^{b}}\right)
\nonumber \\ & = &
\frac{\gamma}{N}\left(\frac{\partial}{\partial t}
- V^{b}\, \frac{\partial}{\partial x^{b}} -
V^{\rm r}\,\frac{\partial}{\partial r} + V^{\rm
r}\,\frac{\partial}{\partial r} - V^{\rm r}\,
\beta^{b}\,\frac{\partial}{\partial x^{b}}\right) \nonumber
\\ & = & \gamma\, e_{\bot} + v\gamma\,
e_{\vdash}\, .
\end{eqnarray}
The complete set of transformations is
\begin{eqnarray}
\begin{array}{rcl} e_{\bot '} & = & \gamma\, e_{\bot} + v\gamma\,
e_{\vdash} \\
e_{\vdash '} & = & v\gamma\, e_{\bot} + \gamma\, e_{\vdash}\\
e_{\hat{a}} & = & e_{\hat{a}} \end{array} & \hspace{1cm} &
\begin{array}{rcl} e^{\bot '} & = & \gamma\, e^{\bot}
- v\gamma\, e^{\vdash} \\
e^{\vdash '} & = & - v\gamma\, e^{\bot} + \gamma\,
e^{\vdash} \\ e^{\hat{a}} & = & e^{\hat{a}} \end{array}\,\, .
\label{kiwi}
\end{eqnarray}
Notice that the $B$ legs of both the tetrads are the same,
which is why the notation
can be compressed so that TR-gauge tetrad indices like
$\hat{\rho}'$ run over
$(\bot', \hat{a} , \vdash')$.

The inhomogeneous transformation rule describing the behavior
of the spacetime connection coefficients under the above
tetrad transformation is the following:
\begin{equation}
\Gamma^{\hat{\rho}'}\,_{\hat{\sigma}'\hat{\tau}'} =
e^{\hat{\rho}'}\,_{\hat{\sigma}}\,
\Gamma^{\hat{\sigma}}\,_{\hat{\rho}\hat{\mu}}\,
e_{\hat{\sigma}'}\,^{\hat{\rho}}\,
e_{\hat{\tau}'}\,^{\hat{\mu}} +
e^{\hat{\rho}'}\,_{\hat{\sigma}}\,
e_{\hat{\tau}'}\,^{\hat{\mu}}\, e_{\hat{\mu}}
\left[e_{\hat{\sigma}'}\,^{\hat{\sigma}}\right]\, .
\label{banana}
\end{equation}
This law provides the bridge between the TR-gauge connection
coefficients (\ref{TRcc}) and the RT-gauge connection
coefficients (\ref{RTcc}).

\subsection{Geometric link between $\bar{\cal T}$ and $\Sigma$}
As an
example, we apply the developed formalism and derive the
splitting result for the three-boundary extrinsic curvature
$\bar{\Theta}_{ij}$. This result has been obtained via ordinary
tensor methods with projection operators in
Ref.\ \cite{BYL}.
However, the ordinary projection-operator
method is not sufficient
for calculating the analogous split of the $\bar{\cal T}$
Sen connection.
We provide the splitting calculation for
$\bar{\Theta}_{ij}$  here
as a simple demonstration of how such
calculations are performed.
Beginning with the first expression of
(\ref{TRcc}), one uses the
rule (\ref{banana}) in tandem with the set
(\ref{kiwi}) to find
\begin{equation}
\bar{\Theta}^{\hat{r}}\,_{\hat{s}} =
- e^{\hat{r}}\,_{\hat{\sigma}}\,
e_{\hat{s}}\,^{\hat{\mu}}\left( v\gamma\,
\Gamma^{\hat{\sigma}}\,_{\bot\hat{\mu}} +
\gamma\, \Gamma^{\hat{\sigma}}\,_{\vdash\hat{\mu}}\right) -
e^{\hat{r}}\,_{\bot}\,
e_{\hat{s}} \left[ v\gamma\right]
- e^{\hat{r}}\,_{\vdash}\, e_{\hat{s}} \left[
\gamma\right]\, .
\end{equation}
(Note that in this equation $\hat{r}$ and
$\hat{s}$ are $\bar{\cal T}$ triad
indices which take the values
($\bot' , \hat{a}$).) A bit of work
and the relations (\ref{RTcc}) yield the set of
$\bar{\Theta}^{\hat{r}}\,_{\hat{s}}$ triad components,
\begin{eqnarray}
\bar{\Theta}^{\bot'}\,_{\bot'} & = & - \gamma\, a^{\vdash} +
v\gamma\, K^{\vdash}\,_{\vdash}
- \gamma^{3}\, e_{\bot}[v] - v\gamma^{3}\,
e_{\vdash}[v] \nonumber \\
\bar{\Theta}^{\bot'}\,_{\hat{a}} & = &
K^{\vdash}\,_{\hat{a}} - \gamma^{2}\, e_{\hat{a}}[v]
\label{orange} \\
\bar{\Theta}^{\hat{a}}\,_{\hat{c}} & = &
\gamma\, k^{\hat{a}}\,_{\hat{c}} + v\gamma\,
K^{\hat{a}}\,_{\hat{c}}\, .
\nonumber
\end{eqnarray}
With the set (\ref{orange}), construction of the sought-for
splitting of $\bar{\Theta}_{ij}$
is not difficult. For convenience work in spacetime coordinates.
The boundary three-metric may be written as
\begin{equation}\bar{\gamma}_{\mu\nu} = \sigma_{\mu\nu} -
\bar{u}_{\mu}\, \bar{u}_{\nu}\, ,
\end{equation}
where the two-metric $\sigma_{\mu\nu} = g_{\mu\nu}
- \bar{n}_{\mu}\,\bar{n}_{\nu} +
\bar{u}_{\mu}\, \bar{u}_{\nu}$ here serves as the
projection operator into the $B$ slices. Wiring the above form of
$\bar{\gamma}^{\mu}_{\nu}$,
the identity operator on $\bar{\cal T}$,
on each of the free indices of
$\bar{\Theta}_{\mu\nu}$, one obtains
\begin{equation}
\bar{\Theta}_{\mu\nu} =
\bar{u}_{\mu}\, \bar{u}_{\nu}\,
\bar{\Theta}_{\bot'\bot'} -
2\, \bar{u}^{\,}_{(\mu}\,\sigma^{\lambda}_{\nu)}\,
\bar{\Theta}_{\bot'\, \lambda} +
\sigma^{\lambda}_{\mu}\sigma^{\kappa}_{\nu}\,
\bar{\Theta}_{\lambda\kappa}\, , \label{mango}
\end{equation}
where an appeal to the symmetry of $\bar{\Theta}_{\mu\nu}$
has been made. Plugging
$\bar{u}_{\mu} = \gamma\, u_{\mu} + v\gamma\, n_{\mu}$
and the results from (\ref{orange}) into (\ref{mango}),
one arrives at the following split of
the three-boundary extrinsic curvature:\cite{Lau3,BYL}
\begin{eqnarray}
\bar{\Theta}_{\mu\nu} & = & \gamma\,
k_{\mu\nu} + v\gamma\, K_{ij}\,
\sigma^{i}_{\mu}\, \sigma^{j}_{\nu} \nonumber \\
& & + \left\{\gamma^{2}\,
u_{\mu}\, u_{\nu} + 2\, v\gamma^{2}\, u_{(\mu}\,n_{\nu)} +
(v\gamma)^{2}\,
n_{\mu}\, n_{\nu}\right\} \label{beetle} \\
& & \times \left\{ \gamma\,
n^{i}\, a_{i} - v\gamma\, n^{i}\, n^{j}\, K_{ij} +
\gamma^{3}\, u[v] +
v\gamma^{3}\, n[v]\right\} \nonumber \\
& & + 2\, \left\{ \gamma\,
u^{}_{(\mu}\, \sigma^{i}_{\nu)} + v\gamma\, n^{}_{(\mu}\,
\sigma^{i}_{\nu)} \right\} \left\{ n^{j}\, K_{ij} -
\gamma^{2}\, D_{i} v
\right\} \nonumber
\end{eqnarray}
Enforcement of the clamping condition $v \rightarrow
0$ recovers equation (A.16) of Ref.\ \cite{BY},
\begin{equation}
\Theta_{\alpha\beta} =
k_{\alpha\beta} + u_{\alpha}\, u_{\beta}\, n^{i}\, a_{i} +
2\, u^{}_{(\alpha}\, \sigma^{i}_{\beta)}\, n^{j}\, K_{ij}\, .
\label{goldenraisen}
\end{equation}

The set of $\bar{\cal T}$ time-gauge connection coefficients is
$\left\{\bar{\tau}^{\hat{a}}\,_{\bot'\bot'}\, ,\,
\bar{\tau}^{\hat{a}}\,_{\bot'\hat{c}}\, ,\,
\bar{\tau}^{\hat{a}}\,_{\hat{c}\bot'}\,
,\, \bar{\tau}^{\hat{a}}\,_{\hat{c} \hat{b}}\right\}$,
where the first two have been considered in the set (\ref{TRcc}).
The splittings of these expressions are
\begin{eqnarray}
\sigma^{\rho}_{\mu}\,
\bar{a}_{\rho} & = &
\gamma^{2}\,\sigma_{\mu}^{\rho}\, a_{\rho}\, +
(v\gamma)^{2}\, \sigma_{\mu}^{\rho}\, b_{\rho}
\hspace{5mm}
\nonumber
\\ \bar{l}_{\mu\nu} & = & \gamma\,
K_{\tau\rho}\,\sigma^{\tau}_{\mu}\,
\sigma^{\rho}_{\nu} + v\gamma\,
k_{\mu\nu} \nonumber \\
\bar{\tau}_{\hat{a}\hat{c}\bot'} & = & \gamma\,
\Gamma_{\hat{a}\hat{c}\bot} + v\gamma\,
\omega_{\hat{a}\hat{c}\vdash} \label{candy}\\
\bar{\tau}_{\hat{a}\hat{c} b}
& = &
\omega_{\hat{a}\hat{c} b}\, . \nonumber
\end{eqnarray}
Using this set and (\ref{beetle}), one finds the following split
of the time-gauge $\bar{\cal T}$ Sen connection 
(pulled back to $B$) in terms of the
radial-gauge $\Sigma$ Sen connection (pulled back to $B$):
\begin{eqnarray}
\sigma^{\lambda}_{\mu} \bar{{\cal A}}^{\bot'}\,_{\lambda} 
& = &  - \sigma^{\lambda}_{\mu}
\left( A^{\vdash}\,_{\lambda}
+ ({\rm i} \gamma^{2}/\kappa) \nabla_{\lambda} v\right)
\nonumber \\
& &  \label{ant} \\
\sigma^{\lambda}_{\mu} 
\bar{{\cal A}}^{\hat{a}}\,_{\lambda} & = &
- \sigma^{\lambda}_{\mu}\left( v\gamma
A^{\hat{a}}\,_{\lambda}
+ {\rm i} \gamma \epsilon^{\hat{a}\hat{c}}
A_{\hat{c}\lambda}\right)\, .
\nonumber
\end{eqnarray}
Now consider $2 \Gamma^{(+)}\,_{\hat{1}\hat{2}\bot}
= \Gamma_{\hat{1}\hat{2}\bot} - {\rm i} a^{j} n_{j}$, and 
the following clamped results ($v \rightarrow 0$ limit) 
for the splitting:
\begin{eqnarray}
{\cal A}^{\bot}\,_{\mu} & = & - u_{\mu} \left( 2/\kappa \right)
\Gamma^{(+)}\,_{\hat{1}\hat{2}\bot} -
\sigma^{\lambda}_{\mu}\, A^{\vdash}\,_{\lambda}
\nonumber \\
& &   \label{clampedsen} \\
\sigma^{\lambda}_{\mu} {\cal A}^{\hat{a}}\,_{\lambda} & = & -
{\rm i} \sigma^{\lambda}_{\mu}
\epsilon^{\hat{a}\hat{c}} A_{\hat{c}\lambda}\, ,
\nonumber
\end{eqnarray}

To find the splitting of the needed pieces of the radial-gauge
$\bar{\Sigma}$ Sen
connection in terms of the radial-gauge $\Sigma$
Sen connection
and other gauge variables, first find the split of
$\bar{A}^{\hat{r}}\,_{j}$ (pulled back to $B$) in terms of the
$\bar{\cal T}$ foliation variables,
\begin{eqnarray}
\sigma^{\lambda}_{\mu}\bar{A}^{\vdash'}\,_{\lambda} & = &
- \sigma^{\lambda}_{\mu}
\bar{\cal A}^{\bot'}\,_{\lambda} \nonumber \\
& &  \\
\sigma^{\lambda}_{\mu}\bar{A}^{\hat{a}}\,_{\lambda} & = &
 - {\rm i} \sigma^{\lambda}_{\mu}
\epsilon^{\hat{a}\hat{c}}
\bar{\cal A}_{\hat{c}\lambda}\, . \nonumber
\end{eqnarray}
Combination of this result with (\ref{ant}) gives
\begin{eqnarray}
\sigma^{\lambda}_{\mu}\bar{A}^{\vdash'}\,_{\lambda} & = &
\sigma^{\lambda}_{\mu} \left( A^{\vdash}\,_{\lambda}
+ ({\rm i}\gamma^{2}/\kappa) \nabla_{\lambda} v\right)
\nonumber \\
& & \label{sensplit} \\
\sigma^{\lambda}_{\mu}\bar{A}^{\hat{a}}\,_{\lambda} & = &
\sigma^{\lambda}_{\mu}\,
\left(\gamma A^{\hat{a}}\,_{\lambda}
+ {\rm i} v \gamma \epsilon^{\hat{a}\hat{c}}
A_{\hat{c}\lambda}\right)\, .
\nonumber
\end{eqnarray}
The boost relations (\ref{boostlaws}) for
$\bar{\varepsilon}$, $\bar{\jmath}_{\vdash'}$,
and $\bar{\jmath}_{a}$ can be derived with these expressions.
To derive the
boost results for $\bar{s}^{ab}$ and $\bar{t}^{ab}$, one must
use these expressions and also the results
\begin{eqnarray}
2\Gamma^{(+)}\,_{\hat{1}\hat{2}\bot'}
& = & \gamma 2\Gamma^{(+)}\,_{\hat{1}\hat{2}\bot}
- v \gamma \kappa A^{\vdash}\,_{\mu} n^{\mu} 
- {\rm i} \gamma^{2} \bar{u}[v]
\nonumber \\ & & \label{sdsplit} \\
\kappa \bar{A}^{\vdash'}\,_{\mu} \bar{n}^{\mu}
& = & \gamma \kappa A^{\vdash}\,_{\mu} n^{\mu} - v \gamma 2\Gamma^{(+)}\,_{\hat{1}\hat{2}\bot} + 
{\rm i} \gamma^{2} \bar{n}[v] \nonumber
\end{eqnarray}
Note that on the left-hand side the selfdual
coefficients are TR-gauge, while
those on the right-hand side are RT-gauge.

\section{Corner terms in the gravitational action}
This appendix presents a simple tetrad method for analyzing
``sharp-corner" terms in the gravitational action
principle. We show how the corner terms in the action
(\ref{cico}) arise. (Using a 
different method, Hayward has made a systematic study of 
such corner terms.\cite{Hayward}) 
As mentioned, the Goldberg action differs
from the Hilbert action by the pure divergence
\begin{equation}
- \frac{1}{2\kappa}\int_{\cal M}
{\rm d}\left(e^{\hat{\rho}} \wedge \sigma_{\hat{\rho}}\right) =
\frac{1}{\kappa} \int_{\cal M} {\rm d}^{4}x\,
\sqrt{- g}\,\nabla_{\mu}\left(
e^{\hat{\rho} \mu}\, e_{\hat{\sigma}}\,^{\nu}\,
\Gamma^{\hat{\sigma}}\,_{\hat{\rho} \nu}\right)\, .
\end{equation}
To ensure that, upon the use of Stokes' theorem, this
divergence gives the desired ``$TrK$" and ``$Tr\Theta$" terms
on the boundary elements, tie
$e_{\hat{0}}$ to the $u = e_{\bot}$
hypersurface normals on $t'$ and $t''$ and tie  $e_{\hat{3}}$ to
the normal $\bar{n} = e_{\vdash'}$ on $\bar{\cal T}$.
However, if these
gauge conditions are enforced simultaneously, then in general
the tetrad is doubled-valued on the corners $B' = B_{t',r''}$ 
and $B'' = B_{t'',r''}$.
Therefore, in order to both retain the desired ``$TrK$" and
``$Tr\Theta$" terms yet avoid double-valuedness on the corners,
use a limit procedure in which the condition on $e_{\hat{3}}$ is
relaxed in a small neighborhood of the corners. Next, consider
the limit as this neighborhood ``shrinks" to the corners.

The precise procedure is as follows. Suppose that
$e_{\hat{0}}$ does indeed coincide with $u$ on $t'$ and $t''$,
but that on $\bar{\cal T}$ the tetrad has the form
\begin{eqnarray}
e_{\hat{0}} & = &
\psi\,\bar{u} - w \psi\,\bar{n} \nonumber \\
e_{\hat{3}} & = &
\psi\,\bar{n} - w \psi\,\bar{u}\, ,  \label{boost}
\end{eqnarray}
where $\psi \equiv (1 - w^{2})^{-1/2}$. For each
$\delta \in [0,1]$, $w = w(x; \delta)$ is a suitably
continuous and differentiable point-dependent boost velocity
defined on $\bar{\cal T}$. Further, for each $\delta$ assume
that $w(x; \delta) = 0$ except on a ``small" neighborhood
${\cal N}_{\delta}$ of the corners $B'$ and $B''$. For each
$\delta$ the set ${\cal N}_{\delta}$ is not connected, but is
comprised of the disjoint union of two pieces
${\cal N}'_{\delta}$ and ${\cal N}''_{\delta}$. The set
${\cal N}'_{\delta}$ is a ``small" region of $\bar{\cal T}$
which contains $B'$,
and in the limit $\delta \rightarrow 0$ we have that
$({\cal N}'_{\delta} - B') \rightarrow \emptyset$.
Similarly, the set ${\cal N}''_{\delta}$ is a ``small"
region of $\bar{\cal T}$ which contains $B''$, and in the limit
$\delta \rightarrow 0$ we have that
$({\cal N}''_{\delta} - B'')  \rightarrow \emptyset$.
Finally, for each $\delta$ demand that $w(x; \delta) = v(x)$
whenever $x \in B' \bigcup B''$. This
ensures that on the corner two-surfaces
$e_{\hat{0}} = u$ and $e_{\hat{3}} = n$.
Our construction provides us with a family of tetrads
parameterized by $\delta$. By construction the member tetrad
corresponding to each value of $\delta$ is TR-gauge on most of
$\bar{\cal T}$, however, as the corners are approached,
each member is continuously boosted until it is RT-gauge
on the corners (see Figure (\ref{deltafig})). 
Hence, each $\delta$ tetrad is single-valued
on the corners. The idea is to use a
$\delta$ tetrad in our divergence expression and consider
\begin{equation}
\frac{1}{\kappa}\int_{\partial{\cal M}}
{\rm d}^{3}x\,\sqrt{^{_{3}}g}\,
Tr K =
\lim_{\delta \rightarrow 0} \frac{1}{\kappa} \int_{\cal M}
{\rm d}^{4}x\,\sqrt{- g}\,\nabla_{\mu}\left(
e^{\hat{\rho} \mu}\, e_{\hat{\sigma}}\,^{\nu}\,
\Gamma^{\hat{\sigma}}\,_{\hat{\rho} \nu}\right)\, ,
\end{equation}
where the expression on the left-hand side symbolically
represents the integral of the trace of the extrinsic curvature
of $\partial {\cal M}$ as embedded in
${\cal M}$ over all of $\partial {\cal M}$
(which picks up finite corner contributions, since the normal of
$\partial {\cal M}$ changes discontinuously from $u$ to $\bar{n}$
on these two-surfaces). We can use Stokes' theorem to find
\begin{equation}
\frac{1}{\kappa}\int_{\cal M} {\rm d}^{4}x\,\sqrt{- g}\,
\nabla_{\mu}
\left(e^{\hat{\rho} \mu}\, e_{\hat{\sigma}}\,^{\lambda}\,
\Gamma^{\hat{\sigma}}\,_{\hat{\rho} \lambda}\right) =
\frac{1}{\kappa}\int^{t''}_{t'} {\rm d}^{3}x\,\sqrt{h}\, K +
\frac{1}{\kappa}\int_{\bar{\cal T}}
{\rm d}^{3}x\,\sqrt{- \bar{\gamma}}\,
\bar{n}_{\mu}\,
\left(e^{\hat{\rho} \mu}\, e_{\hat{\sigma}}\,^{\lambda}\,
\Gamma^{\hat{\sigma}}\,_{\hat{\rho} \lambda}\right)\, ,
\end{equation}
Focus attention on the $\bar{\cal T}$ boundary term,
\begin{eqnarray}
\lefteqn{\frac{1}{\kappa}\int_{\bar{\cal T}} {\rm d}^{3}x\,
\sqrt{- \bar{\gamma}}\,\bar{n}_{\mu}
\left(e^{\hat{\rho} \mu}\, e_{\hat{\sigma}}\,^{\nu}\,
\Gamma^{\hat{\sigma}}\,_{\hat{\rho} \nu}\right) =} & &
\nonumber \\
& & \hspace{1cm}
\frac{1}{\kappa}\int_{\bar{\cal T}} {\rm d}^{3}x\,
\sqrt{- \bar{\gamma}}\,\bar{n}_{\mu}\,
e^{\hat{\rho} \mu}
\left(e^{\hat{\sigma}}\,_{\hat{\sigma}'}\,
e_{\hat{\rho}}\,^{\hat{\rho}'}\,
\Gamma^{\hat{\sigma}'}\,_{\hat{\rho}' \nu}\,
e_{\hat{\sigma}}\,^{\nu} +
e^{\hat{\sigma}}\,_{\hat{\sigma}'}\, e_{\hat{\sigma}}\!\left[
e_{\hat{\rho}}\,^{\hat{\sigma}'}\right]\right)\, .
\end{eqnarray}
We have used
the inhomogeneous transformation rule for connection
coefficients to express the $\delta$ connection
coefficients in terms of the connection coefficients
$\Gamma^{\hat{\kappa}'}\,_{\hat{\tau}' \nu}$
determined by the TR-gauge tetrad. Using the
the boost relations (\ref{boost}), we find after some
algebra that
\begin{equation}
\frac{1}{\kappa}\int_{\bar{\cal T}} {\rm d}^{3}x\,
\sqrt{- \bar{\gamma}}\,\bar{n}_{\mu}
\left(e^{\hat{\rho} \mu}\, e_{\hat{\sigma}}\,^{\nu}\,
\Gamma^{\hat{\sigma}}\,_{\hat{\rho} \nu}\right) =
- \frac{1}{\kappa}\int_{\bar{\cal T}} {\rm d}^{3}x\,
\sqrt{- \bar{\gamma}}\left(\bar{\Theta} +
\bar{u}[\varphi]\right)\, ,
\end{equation}
where
$\bar{\Theta} =
- \Gamma^{\hat{\sigma}'}\,_{\vdash'\hat{\sigma}'} =
- \Gamma^{\hat{s}}\,_{\vdash'\hat{s}}$
($\hat{s}$ runs over $(\bot',\hat{1},\hat{2})$) and
$\varphi = \varphi(x; \delta) = \tanh^{-1}(w(x; \delta))$.
Next, since
$\bar{u} = 1/\bar{N}(\partial/\partial t - \bar{V})$,
with some integrations by parts the final integral on
the right-hand side becomes
\begin{equation}
- \frac{1}{\kappa}\int_{\bar{\cal T}} {\rm d}^{3}x\,
\sqrt{- \bar{\gamma}}\,\bar{u}[\varphi] =
- \frac{1}{\kappa}\int_{B'}^{B''}
{\rm d}^{2}x\,\sqrt{\sigma}\, \varphi
+ \frac{1}{\kappa}\int_{\bar{\cal T}}
{\rm d}^{3}x\, \varphi\,
\dot{\sqrt{\sigma}\,\,\,\,\,}\!\!\!\!\!
- \frac{1}{\kappa}\int_{\bar{\cal T}}
{\rm d}^{3}x\,\sqrt{\sigma}\, \varphi\,
\left( \delta_{b}\, \bar{V}^{b}
\right)\, .\nonumber
\end{equation}
We have that
$\lim_{\delta \rightarrow 0}\varphi(x;\delta) = 0$
everywhere on $\bar{\cal T}$
except for corner points where
$\lim_{\delta \rightarrow 0}\varphi(x; \delta) = \phi(x)$.
Therefore, in this limit only the first corner-term
integrals on the right-hand side survive.
Hence we have the main result
\begin{equation}
\frac{1}{\kappa}\int_{\partial{\cal M}}
{\rm d}^{3}x\,\sqrt{^{_{3}}g}\, Tr K
= \frac{1}{\kappa}\int^{t''}_{t'}{\rm d}^{3}x\,\sqrt{h}\, K
- \frac{1}{\kappa}\int_{\bar{\cal T}}
{\rm d}^{3}x\,\sqrt{-\bar{\gamma}}\,
\bar{\Theta} - \frac{1}{\kappa}\int_{B'}^{B''}
{\rm d}^{2}x\,\sqrt{\sigma}\,\phi\, ,
\end{equation}
which justifies (\ref{cico}).
Since the action $S^{\it 1}$ in (\ref{cico}) is
essentially a metric action, we have borrowed the
results from Ref.\ \cite{BYL} to obtain the variation
(\ref{parthenon}). However, it is not
difficult to use the $\delta$ tetrad method to obtain this
result. To perform this calculation it helps to assume that
$\delta w = 0$, or, in other words, the
variations of the $\delta$
tetrad and TR-gauge tetrad are ``locked" together. We note
in passing that a straightforward though
somewhat lengthy calculation shows that variation of the
action (\ref{gaction}) is
\begin{eqnarray}
\delta S^{\it 1} & = & -
\frac{1}{\kappa}\int_{\cal M}{\rm d}^{4}x\,\sqrt{-g}\,
G^{\mu\nu}\,e_{\hat{\rho} \mu}\,\delta e^{\hat{\rho}}\,_{\nu}
\nonumber \\
& &
- \frac{1}{\kappa}\,\int_{\cal M}
{\rm d}^{4}x\,\sqrt{-g}\,\nabla_{\mu}
\left[\left(2 \Gamma^{\hat{\rho}\hat{\sigma}}\,_{\hat{\sigma}}\,
e_{\hat{\rho}}\,^{[\mu}\,
e_{\hat{\tau}}\,^{\nu]} -
\Gamma^{\hat{\rho}\hat{\sigma}}\,_{\hat{\tau}}\,
e_{\hat{\rho}}\,^{\mu}\,
e_{\hat{\sigma}}\,^{\nu}\right)
\delta e^{\hat{\tau}}\,_{\nu}\right]\, .
\end{eqnarray}
One must insert the $\delta$ tetrad into this expression
and then take the
limit $\delta \rightarrow 0$.

The pure imaginary boundary term (\ref{dahi}) added to the
Goldberg action may also be
expressed as
\begin{equation}
- S^{\it 0} = \frac{\rm i}{2\kappa}\int_{\cal M}{\rm d}^{4}x\,
\sqrt{-g}\, \nabla_{\sigma}\left(e_{\hat{\sigma}}\,^{\sigma}\,
\epsilon^{\hat{\sigma}\hat{\rho}\hat{\mu}\hat{\tau}}\,
\Gamma_{\hat{\rho}\hat{\mu}\hat{\tau}}\right)\, . \label{sterm}
\end{equation}
The variation of this expression is
\begin{eqnarray}
- \delta S^{\it 0} & = & - \frac{\rm i}{2\kappa}\int_{\cal M}
{\rm d}^{4}x\,\sqrt{-g}\,
\nabla_{\sigma}\,\nabla_{\lambda}
\left(\epsilon^{\sigma\lambda\tau\mu}\,
e_{\hat{\rho} \tau}\, \delta e^{\hat{\rho}}\,_{\mu}\right)
\label{sterm2} \\
& & + \frac{\rm i}{2\kappa}\int_{\cal M}
{\rm d}^{4}x\,\sqrt{-g}\,
\nabla_{\sigma}
\left[\epsilon^{\hat{\sigma}\hat{\rho}\hat{\mu}\hat{\tau}}\,
\Gamma_{\hat{\rho}\hat{\mu}\lambda}
\left(e_{\hat{\tau}}\,^{\lambda}\,
\delta e_{\hat{\sigma}}\,^{\sigma}
+ e_{\hat{\sigma}}\,^{\sigma}\,
\delta e_{\hat{\tau}}\,^{\lambda}
- e_{\hat{\sigma}}\,^{\sigma}\,
 e_{\hat{\tau}}\,^{\lambda}\,
e^{\hat{\kappa}}\,_{\kappa}\,
\delta e_{\hat{\kappa}}\,^{\kappa}\right) \right]
\nonumber
\end{eqnarray}
Using the $\delta$ tetrad in each of the
above expressions, one can take the
$\lim_{\delta \rightarrow 0}$ and verify
that $- S^{\it 0}$ and $- \delta S^{\it 0}$ contribute no
corner terms.


\begin{figure}
\epsfxsize=4in
\centerline{\epsfbox{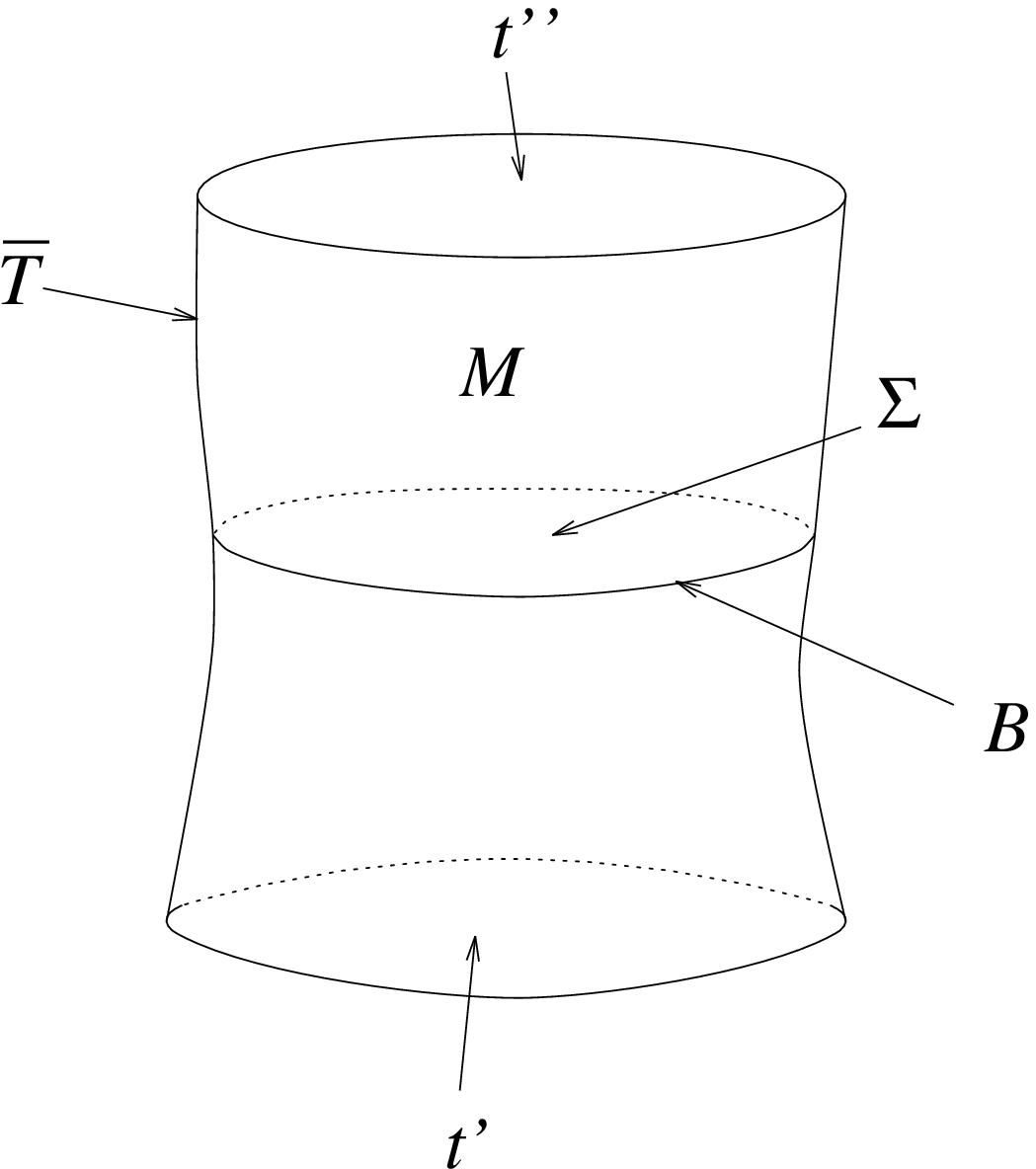}}
\caption{Spatially bounded spacetime region $\cal M$.}
{\small In this simple diagram one dimension is suppressed. 
Although the timelike portion $\bar{\cal T}$ of the boundary of $\cal M$ 
is depicted as connected, this is not a necessary 
restriction.}
\label{spacetimefig}
\end{figure}

\begin{figure}
\epsfxsize=4in
\centerline{\epsfbox{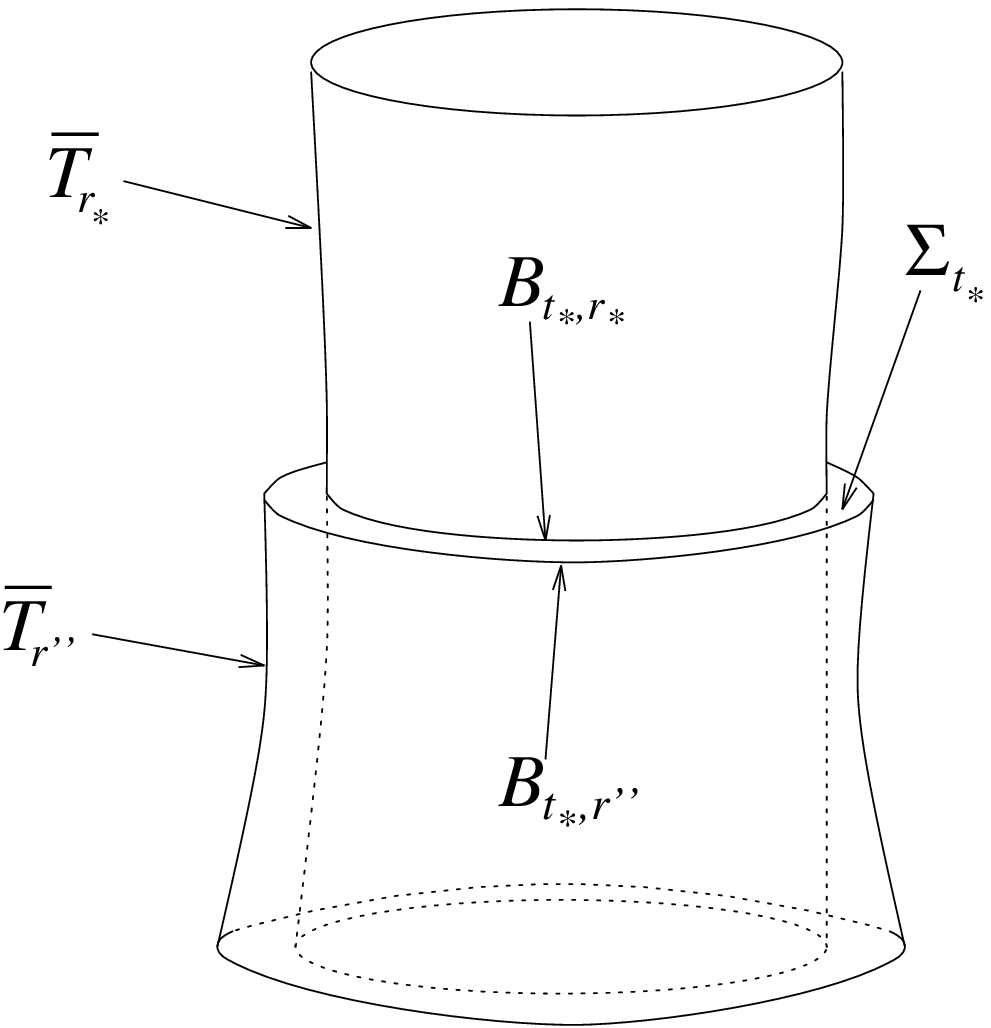}}
\caption{Radial $\bar{\cal T}$ foliation of $\cal M$.}
{\small The diagram depicts a cut-away view of $\cal M$ which shows
the nested timelike sheets of constant $r$. The portion of the
hypersurface $\Sigma_{t_{*}}$ shown lies between the two-surfaces
$B_{t_{*},r_{*}}$ and $B_{t_{*},r''}$. Both $t_{*}$ and $r_{*}$
are constants; $r''$ is the constant which labels the actual boundary 
$\bar{\cal T} = \bar{\cal T}_{r''}$.}
\label{radialfoliationfig}
\end{figure}

\begin{figure}
\epsfxsize=4in
\centerline{\epsfbox{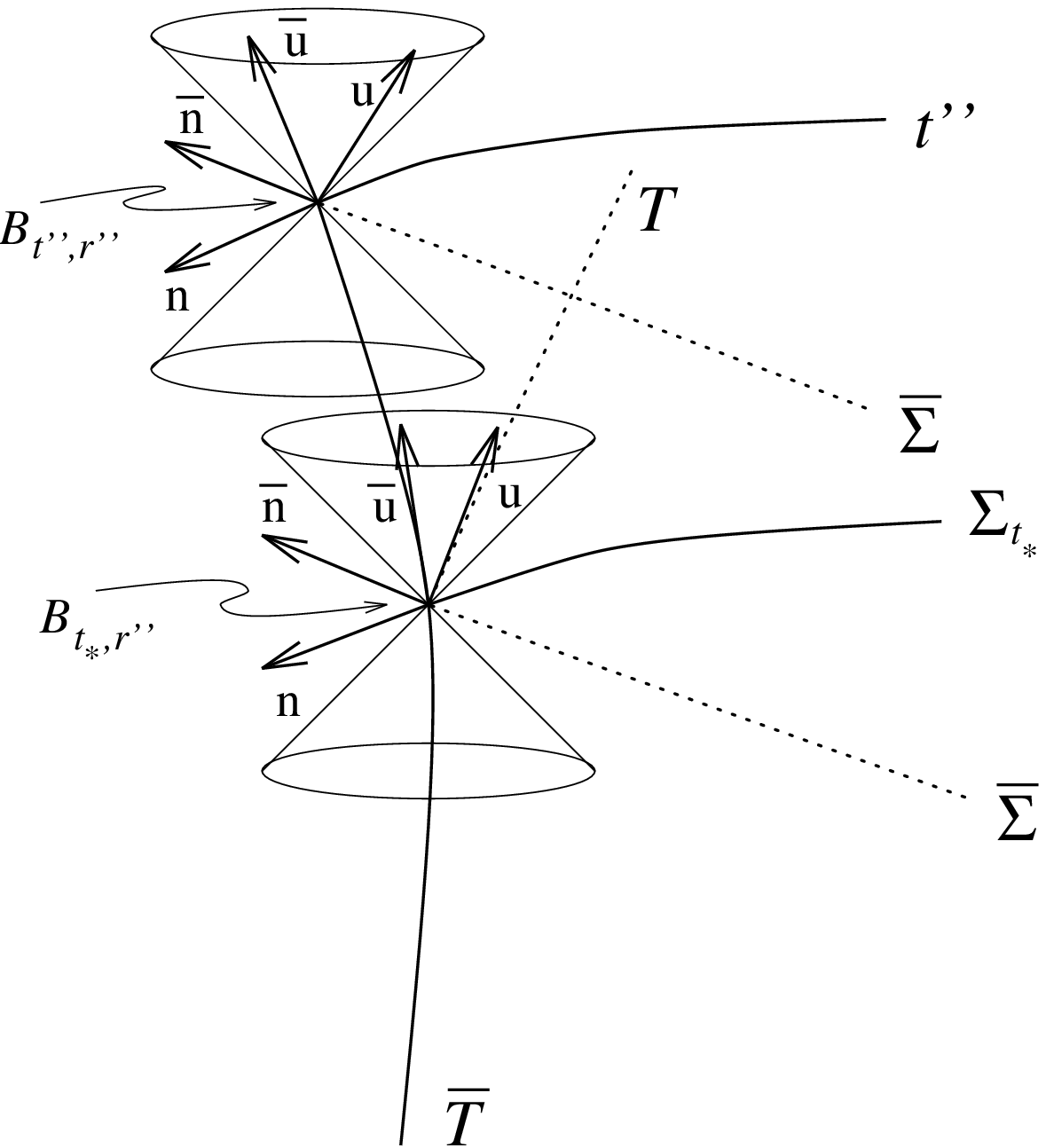}}
\caption{Geometry of an ``unclamped'' $\Sigma$ foliation.}
{\small In this diagram the two-dimensions corresponding to the
two-surfaces $B_{t,r}$ are suppressed (but only one dimension 
of the rather symbolic local lightcones is suppressed). 
The $t = t_{*}$ slice as well as the final 
$t''$ slice of the temporal $\Sigma$ foliation are depicted 
as heavy lines. The actual boundary $\bar{\cal T}$ is also shown 
as a heavy line. Notice that the $\Sigma$ Eulerian history 
(along the integral curves of $u$) of a two-surface $B$ 
generates a timelike ``surface'' $\cal T$ which emerges from (or
crashes into) the actual
boundary. The $\bar{\Sigma}$ slices, depicted with dotted lines,
are orthogonal to the actual boundary, while the $\Sigma$ slices
are orthogonal to the $\cal T$ sheets. When the clamping assumption
is made all $\Sigma$ slices are also $\bar{\Sigma}$ slices, and,
therefore, $\cal T$ and $\bar{\cal T}$ also coincide. When this 
assumption
is made, we drop all bars from the formalism.
RT-gauge tetrads have their time leg locked to $u$ and their
space leg locked to $n$, while TR-gauge tetrads have their 
time leg locked to $\bar{u}$ and their
space leg locked to $\bar{n}$. These vectors are drawn in the
diagram as arrows extending orthogonally
off $B_{t'',r''}$ and $B_{t_{*},r''}$.}
\label{u-and-nfig}
\end{figure}

\begin{figure}
\epsfxsize=4in
\centerline{\epsfbox{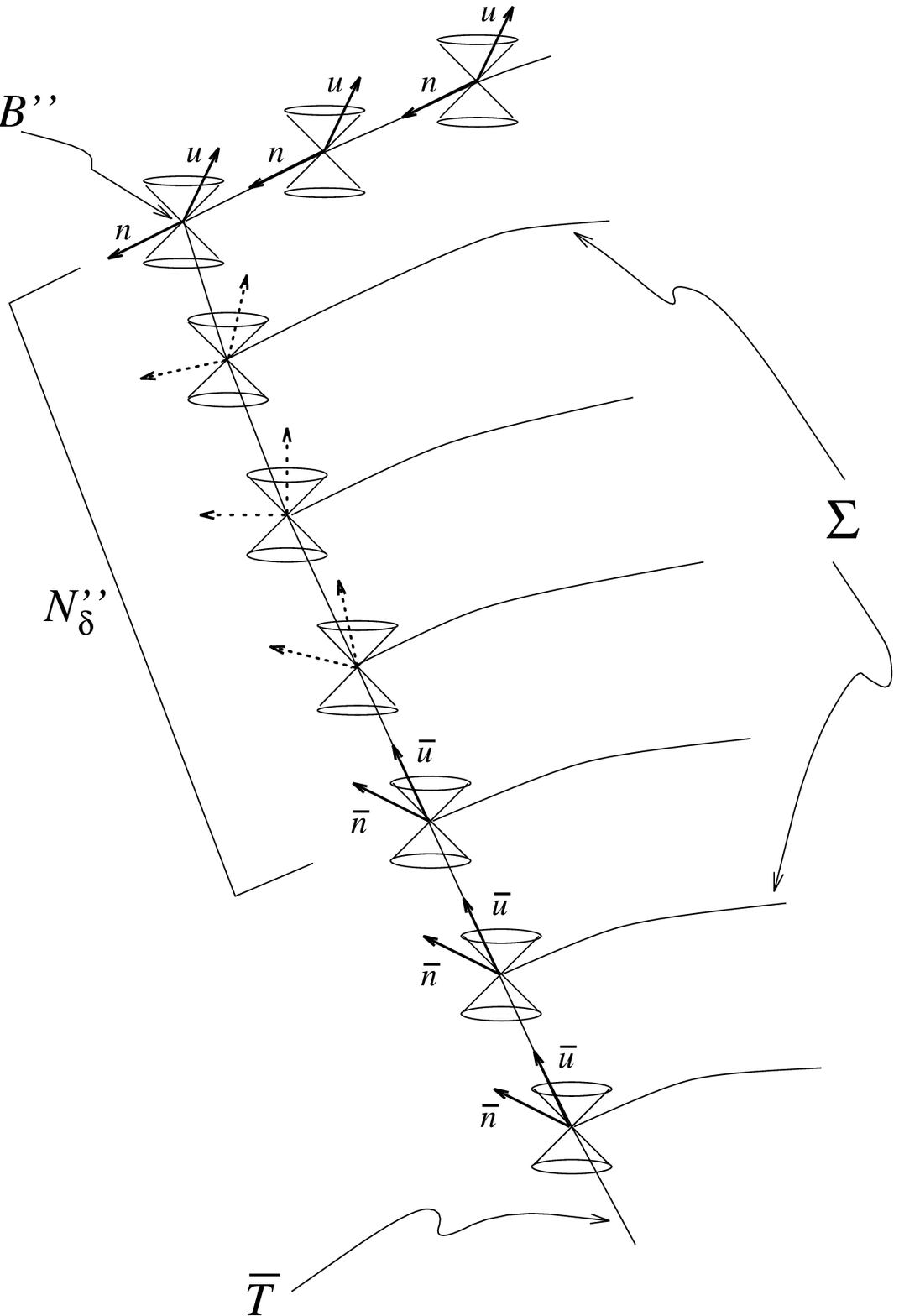}}
\caption{The $\delta$ tetrad.}
{\small In this diagram the two-dimensions 
corresponding to the two-surfaces $B_{t,r}$ are 
suppressed (but only one dimension 
of the rather symbolic local lightcones is suppressed).
The $\delta$ tetrad is TR-gauge on $\bar{\cal T}$ except for
small neighborhoods of the corners, for instance, the neighborhood
${\cal N}''_{\delta}$ of the corner $B_{t'',r''} = B''$. As depicted
in these ``snapshots,''
in such a neighborhood the tetrad is continuously boosted as the
corner is approached and becomes RT-gauge at the corner. 
The dotted vectors drawn around the lightcones on $N''_{\delta}$ are
neither RT gauge nor TR gauge.}
\label{deltafig}
\end{figure}

\end{document}